\documentclass[letterpaper]{JHEP3}
\usepackage{amssymb,amsmath,mathrsfs,slashed}
\usepackage{epsfig}
\usepackage{feynmp}
\usepackage{gastex}

\newcommand{\uD}{\mathrm{D}}
\newcommand{\M}{\mathcal{M}}
\newcommand{\ue}{\mathrm{e}}
\newcommand{\ui}{\mathrm{i}}
\newcommand{\abs}[1]{\left\vert#1\right\vert}
\newcommand{\ud}{\mathrm{d}}
\newcommand{\ve}{\varepsilon}
\newcommand{\SU}[1]{\mathop{\mathrm{SU}}\left(#1\right)}

\newcommand{\SO}[1]{\mathop{\mathrm{SO}}\left(#1\right)}
\newcommand{\KS}{\mathrm{KS}}
\newcommand{\vis}{\mathrm{vis}}

\newcommand{\s}{\mathcal{S}}

\DeclareMathOperator*{\tr}{tr}
\DeclareMathOperator{\csch}{csch}
\DeclareMathOperator{\sech}{sech}

\title{Holographic gauge mediation via strongly coupled messengers}

\author{Paul McGuirk,$^{1}$ Gary Shiu,$^{1,2}$ and Yoske
  Sumitomo$^{1,2}$
  \\
  $^1$Department of Physics, University of Wisconsin, Madison, WI 53706, USA\\
  $^2$Institute for Advanced Study, Hong Kong University of Science
  and
  Technology, Hong Kong, People's Republic of China\\
  \email{mcguirk@physics.wisc.edu, shiu@physics.wisc.edu,
    sumitomo@wisc.edu}}

\date{\today}

\abstract{ We consider a relative of semi-direct gauge mediation where
  the hidden sector exists at large 't~Hooft coupling.  Such scenarios
  can be difficult to describe using perturbative field theory methods
  but may fall into the class of holographic gauge mediation
  scenarios, meaning that they are amenable to the techniques of
  gauge/gravity duality.  We use a recently found gravity solution to
  examine one such case where the hidden sector is a cascading gauge
  theory resulting in a confinement scale not much smaller than the
  messenger mass.  In the original construction of holographic gauge
  mediation, as in other examples of semi-direct gauge mediation at
  strong coupling, the primary contributions to visible sector soft
  terms come from weakly coupled messenger mesons.  In contrast to
  these examples, we describe the dual of a gauge theory where there
  are significant contributions from scales for which the strongly
  coupled messenger quarks are the effective degrees of freedom.  In
  this regime, the visible sector gaugino mass can be calculated
  entirely from holography.  }

\preprint{MAD-TH-09-09}


\begin{document}

\section{Introduction}

A major driving force behind the considerations of physics beyond the
Standard Model (BSM) is arguably the hierarchy problem. Though
countless number of scenarios have been proposed over the past few
decades, they can be broadly divided depending on whether the unknown
physics at the TeV scale is weakly or strongly coupled.  Supersymmetry
(SUSY) is a flagship example of the former.  While the dynamics of a
strongly coupled hidden sector is typically assumed to be the trigger
of SUSY breaking, its influence on the Standard Model and its
supersymmetric extension can be parametrized by a collection of
operators that softly break SUSY. The perturbativity of such weakly
coupled models not only makes them appealing in light of LEP
constraints but also more amenable to quantitative studies. In
comparison, strongly coupled scenarios such as technicolor involve
strong coupling physics at the TeV scale and thus a detailed precision
analysis of such models becomes a highly formidable task.

In supersymmetric scenarios one gains calculability by assuming that
the BSM physics (i.e. the superpartners of the standard model or some
extension) is weakly coupled but the large number of operators that
must be added makes it difficult to make unique predictions (see
e.g.~\cite{Martin:1997ns,Chung:2003fi} for a review).  The situation
can be greatly ameliorated by studying the mechanism by which the
effects of SUSY breaking are mediated to the visible sector.  Of the
different classes of mediation of SUSY breaking, gauge
mediation~\nocite{Dine:1981za,Dimopoulos:1981au,Dine:1981gu,Nappi:1982hm,%
  AlvarezGaume:1981wy,Dimopoulos:1982gm,Dine:1993yw,Dine:1994vc,Dine:1995ag}
\cite{Dine:1981za}-\cite{Dine:1995ag}
(see also~\cite{Giudice:1998bp} for a review and~\cite{Meade:2008wd}
for a very general discussion) has the advantage of suppressing the
flavor-mixing effects that one would generically expect from the
profusion of soft SUSY-breaking operators.  In addition to the visible
sector containing a supersymmetric extension of the standard model,
such models possess fields that can be loosely divided into a hidden
sector, which is neutral under the visible sector gauge group, and a
messenger sector which is charged under the visible sector group.  The
hidden sector, either by design or by assumption obtains a
SUSY-breaking state via strong dynamics (see e.g ~\cite{Shadmi:1999jy}
for a review of dynamical SUSY breaking). The messenger sector, which
couples to the hidden sector, communicates this effect to the visible
sector fields via quantum effects.

Although in models of gauge mediation the messenger sector is often
taken to be neutral with respect to hidden sector group responsible
for the dynamical breaking of supersymmetry (in which case the
coupling between the messengers and hidden sector typically occurs at
the level of the superpotential), it is interesting to consider cases
where this assumption is relaxed.  In models of direct mediation, such
as those
in~\cite{Affleck:1984xz,Poppitz:1996fw,ArkaniHamed:1997jv,Murayama:1997pb},
the distinction between the messengers and hidden sector is less sharp
as the messengers are involved in the dynamical breaking of
supersymmetry.  Between these two extremes is semi-direct gauge
mediation~\cite{Seiberg:2008qj} in which the messengers are charged
under the hidden sector gauge group, as well as the visible sector
gauge group, but do not participate in the SUSY breaking.  When the
messenger sector is weakly coupled, one can use the language and
techniques of perturbative field theory to calculate the effects on
the visible sector.  However, since SUSY breaking is often taken to
occur via strong dynamics, one may wish to consider scenarios in which
the hidden sector has a large 't~Hooft coupling.  In this case the
messengers, when charged under the hidden sector group, are themselves
strongly coupled and other techniques must be used.  In recent years,
our toolbox for handling strongly coupled gauge theories has expanded
dramatically. Duality symmetries, such as Seiberg duality
\cite{Seiberg:1994pq} and gauge/gravity duality
\cite{Maldacena:1997re,Gubser:1998bc,Witten:1998qj}, have enabled us
to map strong coupling physics to their more tangible weak coupling
duals.  Armed with these tools, we can now explore new BSM scenarios
and/or regions of model spaces which were previously overlooked or
ignored because of the complications with strong coupling.

In this paper, we report on some progress in this direction by
computing holographically the effects of semi-direct gauge mediated
supersymmetry breaking with {\it strongly coupled} messengers.  Though
these models are weakly coupled in the sense that the effects of SUSY
breaking on the visible sector can be expressed in terms of a soft
SUSY-breaking Lagrangian, the fact that the messengers are strongly
coupled with respect to the hidden sector gauge group suggests that
their contributions to soft terms are subject to large hidden sector
loop corrections.  As a result, the way that the messenger mass and
SUSY-breaking scale appear in the soft SUSY Langrangian may differ
from the usual perturbative expressions (e.g.~\eqref{eq:beninimass}
and Fig.~\ref{fig:mesonloop}) which assume weakly coupled messengers.
Fortunately, holographic techniques become useful when the hidden
sector gauge group has large 't~Hooft coupling\footnote{Holographic
  techniques may also be useful for describing strong coupling
  dynamics in the observable sector; see, e.g.~\cite{Kachru:2009kg}
  for recent efforts toward describing a holographic and string
  theoretic embedding of technicolor.}.  In examples where the
holographic dual is known, a tree-level computation on the gravity
side amounts to summing up all loop planar diagrams involving the
messengers and the strongly coupled hidden sector
(Fig.~\ref{fig:quarkloop}).

Our work is motivated by a related interesting scenario suggested in
\cite{Benini:2009ff} and the gravity duals of SUSY-breaking large rank
$\SU{N+M}\times\SU{N}$ gauge theories recently obtained in
\cite{MSS1}.  Utilizing the gravity background presented in
\cite{DeWolfe:2008zy}, the authors of \cite{Benini:2009ff} constructed
the holographic dual of a semi-direct gauge mediation scenario where
the masses of the messenger quarks are much higher than the hidden
sector confinement scale.  The large residual $R$-symmetry preserved
at high energies by the SUSY-breaking state suggests that
contributions to the gaugino mass might be suppressed and indeed to
leading order in the SUSY-breaking order parameter there is no
contribution to the gaugino mass from physics at energies scales above
the messenger quark mass in that scenario.  Rather, the contributions
to the gaugino mass come from below this scale where the effective
degrees of freedom of the messenger sector are the \textit{weakly}
coupled mesonic bound states of the messenger quarks whose
interactions are suppressed by the large 't~Hooft
coupling\footnote{This is similar to~\cite{Izawa:2008ef} where
  composite states of analogous messenger fields were treated as the
  primary source of mediation.}.  Therefore, even though the mesonic
spectrum and effective $F$-terms require a holographic computation
because the hidden sector is strongly coupled, the impact on the MSSM
sector can be described by the usual perturbative expression. In
contrast, in this paper we use the supergravity solutions obtained in
\cite{MSS1} which allow us to work in a different kinematic regime
where the messenger masses are comparable to the confinement scale.
In this regime, we find a contribution from scales even above the
messenger quark mass where the propagating degrees of freedom include
\textit{strongly coupled} quarks.

This paper is organized as follows.  In Section~\ref{sec:holoreview},
we review the holographic approach to gauge mediation suggested
in~\cite{Benini:2009ff}.  Generic arguments based on $R$-symmetry
motivate us to consider one of the non-supersymmetric solutions
presented in~\cite{MSS1} which we briefly summarize in
Section~\ref{subsec:gravsol}.  In Section~\ref{sec:gaugino}, we
compute the visible sector gaugino mass by considering gauginos living
on a stack of probe $\uD 7$-branes in this geometry.  We end with some
discussion in Section \ref{sec:discussion}.  Some useful details about
the deformed conifold geometry and our conventions are relegated to
the appendices.

\section{Holographic gauge mediation}
\label{sec:holoreview}

As in~\cite{Benini:2009ff}, we take the hidden sector at short
distances to be an $\mathcal{N}=1$ $\SU{N+M}\times\SU{N}$ gauge theory
with large `t Hooft couplings.  The matter content of the hidden
sector possesses an $\SU{2}\times\SU{2}$ flavor symmetry under which
the bifundamental chiral multiplets $A_{i=1,2}$ and $B_{i=1,2}$
transform as $\left(\mathbf{2},\mathbf{1}\right)$ and
$\left(\mathbf{1},\mathbf{2}\right)$ respectively.  The superpotential
for this chiral matter is
\begin{equation}
  W_{\mathrm{hidden}}=\lambda_{1}\epsilon^{ij}\epsilon^{kl}
  \tr\left(A_{i}B_{k}A_{j}B_{l}\right).
\end{equation}
For non-vanishing $M$, as the theory flows to the IR, it undergoes a
cascade of Seiberg dualities~\cite{Seiberg:1994pq} ending with an
$\SU{M}$ gauge theory exhibiting confinement at a scale
$\Lambda_{\ve}$.  At short distances, the theory possesses a
$\mathbb{Z}_{2M}$ $R$-symmetry that, in the IR, is spontaneously broken
to $\mathbb{Z}_{2}$ by hidden sector gluino condensation.

At large 't~Hooft coupling, the theory is strongly coupled and can be
difficult to analyze.  However, it is in this limit that the
techniques of the gauge-gravity correspondence can be most reliably
applied.  The gravity dual for the high energy theory (the KT
solution~\cite{Klebanov:2000nc}) was constructed in IIB string
theory\footnote{Our conventions are spelled out in
  Appendix~\ref{app:conv}.} by placing $N$ $\uD 3$-branes and $M$
fractional $\uD 3$-branes (i.e. $\uD 5$-branes that wrap collapsing
two-cycles) at a conifold singularity with the world-volumes filling
the four external spacetime dimensions.  The failure of the KT
solution to describe the IR behavior of the field theory is related to
the presence of the naked conifold singularity in the geometry.  The
KS solution~\cite{Klebanov:2000hb} provides the IR resolution by
smearing $M$ fractional $\uD 3$-branes over the finite $S^{3}$ at the
tip of the deformed conifold.

Although dynamical SUSY breaking in this theory is difficult to
describe using standard field theory techniques, a holographic
realization of a SUSY-breaking state can be constructed by adding
$\overline{\uD 3}$-branes to the tip.  In the absence of $\uD
3$-branes and as long as $P$, the number of $\overline{\uD 3}$-branes,
is much smaller than the amount of flux, the $\overline{\uD 3}$-branes
are perturbatively stable but will quantum mechanically tunnel into a
SUSY-preserving vacuum~\cite{Kachru:2002gs}.  The back reaction of the
$\overline{\uD 3}$-branes (the DKM solution) was found
in~\cite{DeWolfe:2008zy} for the KT region (i.e. at large radius).
The presence of the $\overline{\uD 3}$-branes in the geometry
explicitly breaks SUSY on the gravity side but the rapid fall-off of
the resulting non-SUSY perturbations to the bulk fields due to the
$\overline{\uD 3}$-brane indicates that this configuration is dual to
a metastable SUSY-breaking state, rather than dual to a theory that
explicitly breaks SUSY.

The DKM solution was used in~\cite{Benini:2009ff} to provide a
holographic realization of a scenario of gauge mediation.  The
standard model gauge group is taken as a subgroup of a global $\SU{K}$
symmetry that is introduced into the field theory dual by adding a
stack of $K$ probe $\uD 7$-branes into the geometry and weakly gauged
by adding a cutoff to the geometry (Fig.~\ref{fig:throat}).  These
$\uD 7$-branes fill the large four dimensional spacetime and extend
along the radial direction of the conifold while wrapping a
three-cycle in the angular directions.  The matter content of the
standard model is placed at the cutoff of the geometry, which on the
gauge theory side corresponds to taking the standard model fields
to be elementary fields, rather than composites resulting from the
strong dynamics of the hidden sector.

\EPSFIGURE[t]{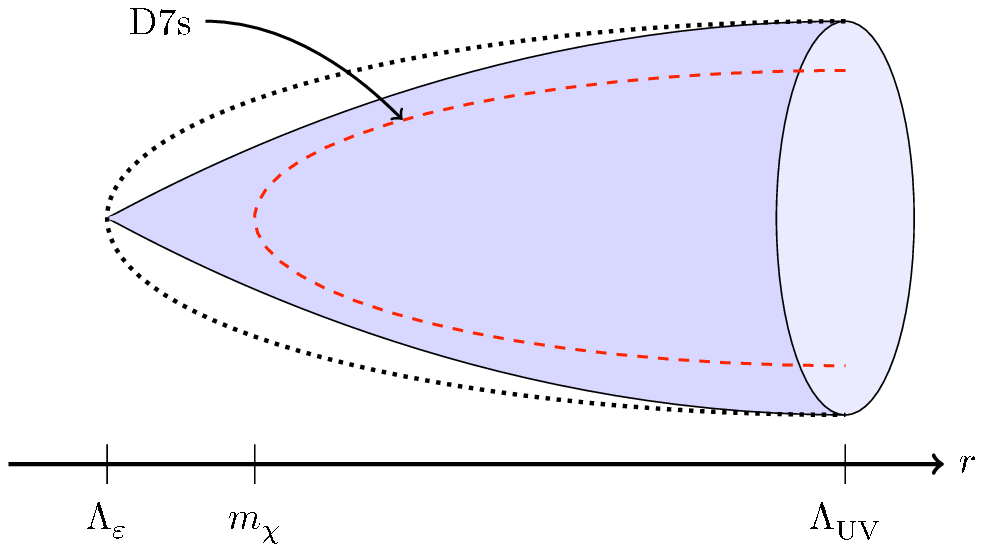}{\label{fig:throat} The flavor branes dip
  down to a distance set by the messenger mass $r=m_{\chi}$ where the
  coordinate $r$ is defined in~\eqref{eq:radialcoordinates} and has
  mass dimension $1$.  Although we will consider a case where
  $m_{\chi}$ is not much larger than $\Lambda_{\ve}$, because of the
  strong warping this corresponds to a large proper distance.  The
  back reaction of the $\overline{\uD 3}$-branes reintroduces a
  singularity at $r=\Lambda_{\ve}$ so that unlike the KS solution
  (represented by the dotted line), it is no longer smooth at the
  tip.}

In particular, the (deformed) conifold inherits a complex structure through the
defining equation
\begin{equation}
  \label{eq:deformconifold}
  \sum_{i=1}^{4}z_{i}^{2}=\ve^{2},\qquad z_{i}\in\mathbb{C},
\end{equation}
where the deformation parameter is related to the confining scale of
the dual gauge theory by $\Lambda_{\ve}=\ve^{2/3}$.  In terms of these
holomorphic coordinates, we take the world-volume of the $\uD 7$-branes
to be specified by the condition
\begin{equation}
  \label{eq:kupersteincondition}
  z_{4}=\mu.
\end{equation}
The addition of $K$ such $\uD 7$-branes corresponds to the addition of the
global $\SU{K}$ flavor symmetry~\cite{Karch:2002sh} and matter fields
$\chi$ and $\tilde{\chi}$ to the gauge theory with $\chi$ transforming
as an anti-fundamental of the $\SU{N}$ factor of the gauge theory and a
fundamental of $\SU{K}$ and $\tilde{\chi}$ as the conjugate
representations~(Fig. \ref{fig:quiver}).  The bifundamental fields $A_{i}$ and
$B_{i}$ couple
to the quarks through the
superpotential~\cite{Ouyang:2003df,Kuperstein:2004hy}
\begin{equation}
  W_{\mathrm{mess}}=
  \tilde{\chi}^{a}\left(A_{1}B_{1}+A_{2}B_{2}-\mu\right)\chi_{a}
  +\lambda_{2}\tilde{\chi}\chi\tilde{\chi}\chi.
\end{equation}
The fields $\chi$, which have mass dimension $\frac{3}{4}$, have mass
$m_{\chi}=\mu^{2/3}$. This choice of embedding of the $\uD 7$
world-volumes is made since although any holomorphically embedded $\uD
7$ is supersymmetric in the (deformed) conifold, $\uD 7$-branes
satisfying a condition other than~(\ref{eq:kupersteincondition})
typically require the existence of non-trivial world-volume flux to
preserve the same supersymmetry as the KS solution~\cite{Chen:2008jj}.

\FIGURE[t]{
  \begin{picture}(120,40)(-25,-20)
    \gasset{AHangle=30,AHLength=3,AHlength=0}
    \node[Nh=15,Nw=15,Nmr=15](A)(10,0){$N+M$}
    \node[Nh=15,Nw=15,Nmr=15](B)(40,0){$N$}
    \node[Nh=15,Nw=15,Nmr=0](C)(70,0){$K$}
    \drawedge[curvedepth=8,AHnb=2](A,B){$A_{1,2}$}
    \drawedge[curvedepth=8,AHnb=2](B,A){$B_{1,2}$}
    \drawedge[curvedepth=8,AHnb=1](B,C){$\chi$}
    \drawedge[curvedepth=8,AHnb=1](C,B){$\tilde{\chi}$}
  \end{picture}
  \caption{\label{fig:quiver}Quiver for the high energy theory.  The
    standard model gauge group is a subgroup of the global $\SU{K}$.}
}

Since $\chi$ and $\tilde{\chi}$ are charged under both the hidden
sector and the visible sector, they are natural candidates for the
messengers of the effect of SUSY breaking.  Additionally, the
SUSY-breaking state exists independently of the presence of the $\uD
7$-branes implying that these messengers do not actively participate in
the dynamical breaking of supersymmetry\footnote{This is at least true
  when the rank of the global symmetry group $K$ is much smaller than
  that of the gauge symmetry in which case the back reaction of the
  $\uD 7$s on the geometry can be neglected.}.  Thus, the setup is
closely related to semi-direct gauge mediation~\cite{Seiberg:2008qj},
although $\chi$ and $\tilde{\chi}$ do have additional superpotential
couplings to the hidden sector chiral matter.  Because the messenger
quarks are charged under the large 't~Hooft hidden sector, there are
potential contributions to the visible sector gaugino mass from all
planar diagrams (Fig.~\ref{fig:quarkloop}).

\FIGURE[t]{
  \begin{fmffile}{quarkdiagram1}
    \begin{fmfgraph*}(40,25) \fmfpen{0.5thin} \fmfleft{l1}
      \fmfright{l2} \fmf{vanilla,label=$\lambda$}{l1,v1}
      \fmf{vanilla,right,tension=.2,label=$\Psi_{\chi}$}{v1,v2}
      \fmf{dashes,left,tension=.2,label=$\chi$}{v1,v2}
      \fmf{vanilla,label=$\lambda$}{v2,l2} \fmfdot{v1,v2} \fmffreeze
      \fmf{boson}{l1,v1} \fmf{boson}{v2,l2}
    \end{fmfgraph*}
  \end{fmffile}
  \begin{fmffile}{quarkdiagram2}
    \begin{fmfgraph}(40,25) \fmfpen{0.5thin} \fmfset{curly_len}{2mm}
      \fmfleft{l1} \fmfright{l2} \fmf{plain}{l1,v1}
      \fmf{plain,right,tension=.2,tag=1}{v1,v2}
      \fmf{dashes,left,tension=.2,tag=2}{v1,v2} \fmf{plain}{v2,l2}
      \fmfdot{v1} \fmfdot{v2} \fmffreeze \fmfipath{p[]}
      \fmfiset{p1}{vpath1(__v1,__v2)} \fmfiset{p2}{vpath2(__v1,__v2)}
      \fmfi{curly}{point length(p1)/2 of p1 -- point length(p2)/2 of
        p2} \fmfiv{d.sh=circle,d.f=1,d.siz=2thick}{point length(p1)/2
        of p1} \fmfiv{d.sh=circle,d.f=1,d.siz=2thick}{point
        length(p2)/2 of p2} \fmf{boson}{l1,v1} \fmf{boson}{v2,l2}
    \end{fmfgraph}
  \end{fmffile}
  \begin{fmffile}{quarkdiagram3}
    \begin{fmfgraph}(40,25) \fmfpen{0.5thin} \fmfset{curly_len}{2mm}
      \fmfleft{l1} \fmfright{l2} \fmf{plain}{l1,v1}
      \fmf{plain,right,tension=.2,tag=1}{v1,v2}
      \fmf{dashes,left,tension=.2,tag=2}{v1,v2} \fmf{plain}{v2,l2}
      \fmfdot{v1} \fmfdot{v2} \fmffreeze \fmfipath{p[]}
      \fmfipair{vert[]} \fmfiset{p1}{vpath1(__v1,__v2)}
      \fmfiset{p2}{vpath2(__v1,__v2)} \fmfiset{p3}{point length(p1)/2
        of p1 -- point length(p2)/2 of p2} \fmfiset{vert1}{point
        length(p3)/3 of p3} \fmfiset{vert2}{point 2*length(p3)/3 of
        p3} \fmffreeze \fmfi{curly}{subpath (0, length(p3)/3) of p3}
      \fmfi{curly}{subpath (2*length(p3)/3, length(p3)) of p3}
      \fmfi{curly}{vert1{right} .. vert2} \fmfi{curly}{vert2{left}
        .. vert1} \fmfiv{d.sh=circle,d.f=1,d.siz=2thick}{vert1}
      \fmfiv{d.sh=circle,d.f=1,d.siz=2thick}{vert2}
      \fmfiv{d.sh=circle,d.f=1,d.siz=2thick}{point length(p1)/2 of p1}
      \fmfiv{d.sh=circle,d.f=1,d.siz=2thick}{point length(p2)/2 of p2}
      \fmf{boson}{l1,v1} \fmf{boson}{v2,l2}
    \end{fmfgraph}
  \end{fmffile}
  \begin{fmffile}{quarkdiagram4}
    \begin{fmfgraph}(40,25) \fmfpen{0.5thin} \fmfset{curly_len}{2mm}
      \fmfleft{l1} \fmfright{l2} \fmf{plain}{l1,v1}
      \fmf{plain,right,tension=.15,tag=1}{v1,v2}
      \fmf{dashes,left,tension=.15,tag=2}{v1,v2} \fmf{plain}{v2,l2}
      \fmfdot{v1} \fmfdot{v2} \fmffreeze \fmfipath{p[]}
      \fmfipair{vert[]} \fmfiset{p1}{vpath1(__v1,__v2)}
      \fmfiset{p2}{vpath2(__v1,__v2)} \fmfiset{p3}{point length(p1)/2
        of p1 -- point length(p2)/2 of p2} \fmfiset{vert1}{point
        length(p3)/3 of p3} \fmfiset{vert2}{point 2*length(p3)/3 of
        p3} \fmffreeze \fmfi{curly}{point length(p1)/3 of p1 .. vert1}
      \fmfi{curly}{vert1 .. point 2length(p2)/3 of p1}
      \fmfi{curly}{point length(p2)/3 of p2 .. vert2}
      \fmfi{curly}{vert2 .. point 2length(p2)/3 of p2}
      \fmfi{curly}{vert1{right} .. vert2} \fmfi{curly}{vert2{left}
        .. vert1} \fmfiv{d.sh=circle,d.f=1,d.siz=2thick}{vert1}
      \fmfiv{d.sh=circle,d.f=1,d.siz=2thick}{vert2}
      \fmfiv{d.sh=circle,d.f=1,d.siz=2thick}{point length(p1)/3 of p1}
      \fmfiv{d.sh=circle,d.f=1,d.siz=2thick}{point 2length(p1)/3 of
        p1} \fmfiv{d.sh=circle,d.f=1,d.siz=2thick}{point length(p2)/3
        of p2} \fmfiv{d.sh=circle,d.f=1,d.siz=2thick}{point
        2length(p2)/3 of p2} \fmf{boson}{l1,v1} \fmf{boson}{v2,l2}
    \end{fmfgraph}
  \end{fmffile}
  \begin{fmffile}{quarkdiagram5}
    \begin{fmfgraph}(40,25) \fmfpen{0.5thin} \fmfset{curly_len}{2mm}
      \fmfleft{l1} \fmfright{l2} \fmf{plain}{l1,v1}
      \fmf{phantom,right,tension=.15,tag=1}{v1,v2}
      \fmf{phantom,left,tension=.15,tag=2}{v1,v2} \fmf{plain}{v2,l2}
      \fmfdot{v1} \fmfdot{v2} \fmffreeze \fmfipath{p[]}
      \fmfipair{vert[]} \fmfiset{p1}{vpath1(__v1,__v2)}
      \fmfiset{p2}{vpath2(__v1,__v2)} \fmfiset{p3}{point length(p1)/2
        of p1 -- point length(p2)/2 of p2} \fmffreeze
      \fmfi{dashes}{subpath (0,length(p2)/2) of p2}
      \fmfi{vanilla}{subpath (length(p2)/2,length(p2)) of p2}
      \fmfi{vanilla}{subpath (0,length(p1)/2) of p1}
      \fmfi{dashes}{subpath (length(p1)/2,length(p1)) of p1}
      \fmfiset{vert1}{point length(p3)/3 of p3} \fmfiset{vert2}{point
        2*length(p3)/3 of p3} \fmfi{curly}{subpath (0, length(p3)/3)
        of p3} \fmfi{vanilla}{subpath (0, length(p3)/3) of p3}
      \fmfi{curly}{subpath (2*length(p3)/3, length(p3)) of p3}
      \fmfi{vanilla}{subpath (2*length(p3)/3, length(p3)) of p3}
      \fmfi{curly}{vert1{right} .. vert2} \fmfi{vanilla}{vert1{right}
        .. vert2} \fmfi{curly}{vert2{left} .. vert1}
      \fmfiv{d.sh=circle,d.f=1,d.siz=2thick}{vert1}
      \fmfiv{d.sh=circle,d.f=1,d.siz=2thick}{vert2}
      \fmfiv{d.sh=circle,d.f=1,d.siz=2thick}{point length(p1)/2 of p1}
      \fmfiv{d.sh=circle,d.f=1,d.siz=2thick}{point length(p2)/2 of p2}
      \fmf{boson}{l1,v1} \fmf{boson}{v2,l2}
    \end{fmfgraph}
  \end{fmffile}
  \caption{\label{fig:quarkloop}A very small sample of the infinite
    number of loops that might contribute to the visible sector
    gaugino mass.  The gaugino couples to the messenger quarks
    $\Psi_{\chi}$ and squarks $\chi$ which also couple to the large
    't~Hooft coupling hidden sector gluons and gluinos.  Since the
    hidden sector has large 't~Hooft coupling, there are leading order
    contributions from planar diagrams with arbitrary numbers of
    loops.  The calculation can be done holographically and, to
    leading order in the SUSY-breaking parameter, the loops cancel for
    $m_{\chi}\gg\Lambda_{\ve}$.  However, for
    $m_{\chi}\approx\Lambda_{\ve}$, the cancellation no longer
    occurs.}
}

Since the analysis of~\cite{Benini:2009ff} was performed at large
radius on the gravity side, the dual field theory is in a regime where
the messengers are much heavier than the confining scale of the
strongly coupled hidden sector, $m_{\chi}\gg \Lambda_{\varepsilon}$,
and for many parts of that analysis, it is appropriate to neglect
$\varepsilon$.  In the absence of the deformation of the conifold
singularity, the $R$-symmetry preserved by the geometry is
$\mathbb{Z}_{2M}$~\cite{Herzog:2001xk,Klebanov:2002gr}.  This large
amount of $R$-symmetry suppresses contributions to the gaugino mass
from scales above the messenger mass\footnote{The gaugino mass
  $m_{1/2}$ carries two units of $R$-charge, implying that any
  $R$-symmetry larger than $\mathbb{Z}_{2}$ forbids a non-vanishing
  $m_{1/2}$.}.  A non-vanishing messenger mass $\mu$, which has unit
$R$-charge, breaks $R$-symmetry altogether.  However, the $R$-symmetry
breaking effect seems to be small and indeed in~\cite{Benini:2009ff}
the messenger quarks $\chi$ do not directly contribute to the gaugino
mass until higher order in perturbation theory. As a result of the
strong dynamics, the messenger $\chi$ fields bind into mesons
$\Phi_{n}$ which are neutral under the hidden sector gauge group but
transform as adjoints under the visible $\SU{K}$.  The spectrum of
mesons includes states whose masses are below $m_{\chi}$ and the
SUSY-breaking dynamics of the hidden sector cause these meson
superfields to feel $R$-symmetry breaking effective $F$-terms which
lead to a non-vanishing gaugino mass.  Because the hidden sector gauge
group has large rank, the mesons are weakly
coupled~\cite{'tHooft:1973jz,'tHooft:1974hx,Witten:1979kh} and their
physics can be described using standard field theory techniques,
though the spectra $\bigl\{M_{n},F_{n}\bigr\}$ of masses and $F$-terms
do require a holographic calculation (which was performed via a DBI
analysis in~\cite{Benini:2009ff}).  The mesons effectively act as
messengers in a minimal gauge mediation scenario
(Fig.~\ref{fig:mesonloop}) and the result of~\cite{Benini:2009ff} is
that the visible sector gauginos receive a mass
\begin{equation}
  \label{eq:beninimass}
  m_{1/2}=\frac{g^{2}_{\mathrm{vis}}K}{16\pi^{2}}\sum_{n}\frac{F_{n}}{M_{n}}
  \sim\frac{g^{2}_{\mathrm{vis}}K}{16\pi^{2}}\frac{\Lambda_{\s}^{4}}{m_{\chi}^{3}
    \sqrt{4\pi\lambda_{\mathrm{eff}}\left(m_{\chi}\right)}}
  \sum_{n}n\, \ue^{\ui\theta_{n}},
\end{equation}
where $g_{\mathrm{vis}}$ is the visible sector gauge coupling,
$\Lambda_{\s}^{4}$ is the exponentially small vacuum energy of the
SUSY-breaking state (which on the gravity side is set by the warped
tension of the $\overline{\uD 3}$-branes),
$\lambda_{\mathrm{eff}}\left(m_{\chi}\right)$ is the 't~Hooft coupling
of the hidden sector (dual to the amount of effective $\uD 3$-charge)
evaluated at the energy scale $m_{\chi}$, and $\theta_{n}$ are
uncalculated phases.  The summation is over a range of $n$ such that
the effective field theory of weakly coupled mesons is appropriate.
The effects of SUSY breaking is communicated to the remaining visible
sector fields via gaugino
mediation~\cite{Kaplan:1999ac,Chacko:1999mi}.

\FIGURE[t]{
  \begin{fmffile}{mesondiagram}
    \begin{fmfgraph*}(90,45) \fmfpen{0.5thin} \fmfleft{l1}
      \fmfright{l2} \fmf{plain}{l1,v1}
      \fmf{dbl_plain,label=$\Psi_{n}$}{v1,v3} \fmf{dbl_plain}{v2,v3}
      \fmf{phantom,left,tension=.2,tag=1}{v1,v2} \fmf{plain}{l2,v2}
      \fmfdot{v1} \fmfdot{v2}
      \fmfv{decor.shape=cross,decor.filled=full,decor.size=5thick,
        label=$M_{n}$,label.angle=-90}{v3} \fmflabel{$\lambda$}{l1}
      \fmflabel{$\lambda$}{l2} \fmfposition \fmfipath{p[]}
      \fmfiset{p1}{vpath1(__v1,__v2)}
      \fmfi{dbl_dashes,label=$\Phi_{n}$}{subpath (0,length(p1)/2) of
        p1} \fmfi{dbl_dashes}{subpath (length(p1)/2,length(p1)) of p1}
      \fmfiv{d.sh=cross,d.siz=5thick,label=$F_{n}$}{point length(p1)/2
        of p1} \fmf{boson}{l1,v1} \fmf{boson}{l2,v2}
    \end{fmfgraph*}
  \end{fmffile}
  \caption{\label{fig:mesonloop}Contribution to the visible sector
    gaugino mass from a messenger mesons $\Phi_{n}$ and the
    superpartner mesinos $\Psi_{n}$.  In the 't~Hooft limit, the mesons
    and mesinos are weakly coupled and this diagram gives the leading
    order contribution from the mesons, giving~\eqref{eq:beninimass}.}
}

\subsection{A non-SUSY deformation of Klebanov-Strassler}
\label{subsec:gravsol}

In the far IR of the field theory, $R$-symmetry is broken down to
$\mathbb{Z}_{2}$ by hidden sector gluino condensation.  One would expect then
that for $m_{\chi}\sim
\Lambda_{\ve}$, there will be contributions to the gaugino
mass even from energies above $m_{\chi}$.  Indeed, it was estimated
in~\cite{Benini:2009ff} that there should be a contribution to the
gaugino mass from a finite deformation given by
\begin{equation}
  \label{eq:beninimass2}
  \delta m_{1/2}\sim \frac{\Lambda_{\ve}}{m_{\chi}}
  \frac{\Lambda_{\s}^{4}}{m_{\chi}^{3}
    \sqrt{4\pi\lambda_{\mathrm{eff}}\left(m_{\chi}\right)}}.
\end{equation}
For $\mu\gg\ve$ where the DKM solution is valid, this is much smaller
than Eq.~(\ref{eq:beninimass}).  In order for this contribution to be
important compared to that of the meson messengers, it is necessary
that\footnote{Note that this only ensures that the contribution is
  comparable to that of a single meson, while many mesons contribute
  to~\eqref{eq:beninimass}.}
\begin{equation}
  \frac{\Lambda_{\ve}}{m_{\chi}}\gtrsim\frac{g_{\mathrm{vis}}^{2}K}{16\pi^{2}}.
\end{equation}
As an estimate, we can suppose that the global symmetry has $K=5$
and forms an $\SU{5}$ GUT with
$\alpha_{\mathrm{GUT}}\sim\frac{1}{25}$. This gives
\begin{equation}
  \label{eq:gutbound}
  m_{\chi}\lesssim 60 \Lambda_{\ve}.
\end{equation}

Clearly, as the hierarchy between $\Lambda_{\ve}$ and $m_{\chi}$ is
reduced, the more important the $R$-breaking effects of confinement
become.  However, there is a possible concern with taking $m_{\chi}$
to be too small.  On the gravity side of the calculation, decreasing
$m_{\chi}$ corresponds to allowing the probe $\uD 7$s to dip further
into the throat, reaching smaller values of $\tau$.  The presence of a
$\overline{\uD 3}$-brane introduces a curvature singularity into the
back-reacted geometry at $\tau=0$~\cite{MSS1}.  Such a singularity
indicates the supergravity approximation of string theory breaks down
and so the solution should be modified at distances below the string
length.  Thus, in order to trust our analysis of the gaugino mass, the
$\uD 7$-branes must not extend too deeply into the throat.  For small
radial distances, the KS metric~(\ref{eq:metric}) takes the
approximate form\footnote{This metric does not quite satisfy the
  equations of motion; in addition to the existence of corrections
  that are higher order in $\tau$, there are also corrections that are
  $\mathcal{O}\left(\tau^{0}\right)$ but are negligible in the limit
  of large $g_{s}M$.}
\begin{equation}
  \ud s_{10}^{2}\approx h_{0}^{-1/2}\eta_{\mu\nu}\ud x^{\mu}\ud x^{\nu}
  + h_{0}^{1/2}\left(\frac{1}{2}\ud\tau^{2}
    +\ud\Omega_{3}^{2}+\frac{1}{4}\tau^{2}
    \left[g_{1}^{2}+g_{2}^{2}\right]\right),
\end{equation}
where $\ud\Omega_{3}^{2}$ is the line element for a unit $S^{3}$,
$g_{i}$ are other angular 1-forms, and
$h_{0}\sim\left(g_{s}M\right)^{2}$.

We can estimate the string length for strings stretching along the
radial direction at small $\tau$ by considering the world-sheet
action,
\begin{align}
  S_{\sigma}=&-\frac{1}{2\pi\alpha'}\int_{\mathcal{M}}\ud^{2}\sigma\,
  \sqrt{-\gamma}\gamma^{ab}g_{MN}\partial_{a}X^{M}\partial_{b}X^{N} \notag \\
  \sim&-\frac{g_{s}M}{2\pi\alpha'}\int_{\mathcal{M}}\ud^{2}\sigma\,
  \sqrt{-\gamma}\gamma^{ab}
  \partial_{a}X^{\tau}\partial_{b}X^{\tau}.
\end{align}
This implies that the effective string length for strings stretching
along the holographic or internal angular directions is
\begin{equation}
  \sim\sqrt{\frac{2\pi\alpha'}{g_{s}M}}.
\end{equation}
The unknown stringy modification of the geometry can be neglected if
the $\uD 7$-branes remain much further than a string length from the
location of the $\overline{\uD 3}$-branes.  Temporarily setting
$2\pi\alpha'=1$, this condition becomes
\begin{equation}
  \label{eq:taulowerbound}
  \tau_{\mathrm{min}} \gg \frac{1}{g_{s}M}.
\end{equation}
In order for the supergravity approximation to be valid away from
the singularity, $g_{s}M$ must be large so the stringy resolution is
important only for very small values of $\tau_{\mathrm{min}}$.

For a world-volume specified by the embedding
condition~(\ref{eq:kupersteincondition}), the $\uD 7$-brane will
extend to a minimum $\tau$ given by~\cite{Kuperstein:2004hy}
\begin{equation}
  \tau_{\mathrm{min}}=2\, \mathrm{arccosh}\frac{\mu}{\ve}.
\end{equation}
Combining~(\ref{eq:gutbound}) with~(\ref{eq:taulowerbound}) and using
the relationships $\Lambda_{\ve}=\ve^{2/3}$ and $m_{\chi}=\mu^{2/3}$,
we get the expectation that there will be important and calculable
contributions to the gaugino mass when the $\uD 7$-branes reach a
minimum value $\tau_{\mathrm{min}}$ satisfying
\begin{equation}
  \frac{1}{g_{s}M}\ll\tau_{\mathrm{min}}\lesssim 14.
\end{equation}
Given the relative complexity of the KS solution itself, an exact
solution corresponding to the addition of an $\overline{\uD 3}$-brane
would be difficult to find.  Instead, we will limit ourselves
to a small $\tau$ expansion and take
\begin{equation}
\label{eq:taurange}
  \frac{1}{g_{s}M}\ll\tau_{\mathrm{min}} < 1.
\end{equation}
In terms of the dual field theory variables, this means that we are
taking the confining scale $\Lambda_{\ve}$ and the messenger mass
$m_{\chi}$ to be very near each other, but still requiring that latter
be slightly larger.  For simplicity, we take both $\ve$ and $\mu$ to
be real.  The final result of our calculation will be a contribution
to the gaugino mass that differs from~(\ref{eq:beninimass2}) (which is
not necessarily a contradiction since the result was obtained in a
regime where $\Lambda_{\ve}/m_{\chi}$ is a good expansion parameter
while it is not for the calculation presented here).
Nevertheless,~(\ref{eq:beninimass2}) provides a good motivation to
consider deformations to KS at small radius especially since our
calculation will yield a contribution that is enhanced by the hidden
sector 't~Hooft coupling relative to the
estimate~\eqref{eq:beninimass2}.

In~\cite{MSS1}, we found small $\tau$ expansions for
non-supersymmetric perturbations to the KS solution.  For a choice of
parameters, one of these solutions corresponds to the addition of $P$
$\uD 3$-$\overline{\uD 3}$ brane pairs smeared over the finite $S^{3}$
at the tip and is the small radius analogue of the DKM
solution~\cite{DeWolfe:2008zy}\footnote{The solution neglects in the
  supergravity limit the stringy annihilation of the pairs as was done
  in~\cite{DeWolfe:2008zy}.}.  In terms of the angular 1-forms
$g_{i}$ (which are reviewed in Appendix~\ref{app:conifold}), the
metric of this solution is of the warped type
\begin{subequations}
  \label{eq:metric}
  \begin{equation}
    \label{eq:warppedmetric}
    \ud s_{10}^{2}=
    h^{-1/2}\left(\tau\right)\eta_{\mu\nu}\ud x^{\mu}\ud x^{\nu}
    + h^{1/2}\left(\tau\right)\ud \tilde{s}_{6}^{2},
  \end{equation}
  where $\mu=0,1,2,3$ and where the radial and internal angular part
  of the metric is
  \begin{equation}
    \label{eq:internalmetric}
    \ud \tilde{s}_{6}^{2}= 
    p\left(\tau\right)\ud\tau^{2}
    + u\left(\tau\right) g_{5}^{2}
    + q\left(\tau\right)\left(g_{3}^{2}+g_{4}^{2}\right)
    + s\left(\tau\right)\left(g_{1}^{2}+g_{2}^{2}\right).
  \end{equation}
\end{subequations}
Similar to what was found in~\cite{DeWolfe:2008zy}, the presence of
the $\overline{\uD 3}$-branes ``squashes'' the unwarped 6D space
so that it is no longer the geometry of the deformed conifold.
Expanding the perturbation to leading order in $\tau$ gives
\begin{align}
  p\left(\tau\right)=&p_{\KS},
  &u\left(\tau\right)=&u_{\KS}\left(1+\frac{u_{0}}{\tau}\right),\notag \\
  q\left(\tau\right)=&q_{\KS}\left(1+\frac{q_{0}}{\tau}\right),
  &s\left(\tau\right)=&s_{\KS}\left(1+\frac{s_{0}}{\tau}\right),
\end{align}
where the Klebanov-Strassler solution has
\begin{align}
  p_{\KS}\left(\tau\right)=&u_{\KS}\left(\tau\right)=\frac{\ve^{4/3}}
  {6K^{2}\left(\tau\right)},\qquad
  q_{\KS}\left(\tau\right)=\frac{\ve^{4/3}}{2}K\left(\tau\right)
  \cosh^{2}\frac{\tau}{2},\notag \\
  s_{\KS}\left(\tau\right)=&\frac{\ve^{4/3}}{2}K\left(\tau\right)
  \sinh^{2}\frac{\tau}{2},
\end{align}
with
\begin{equation}
  K\left(\tau\right)=
  \frac{\left(\sinh 2\tau-2\tau\right)^{1/3}}{2^{1/3}\sinh\tau}.
\end{equation}
The presence of the $\overline{\uD 3}$-branes perturbs the geometry so that
\begin{equation}
  u_{0}\sim q_{0}\sim s_{0}\sim \mathcal{S},
\end{equation}
where $\mathcal{S}$ is proportional to the number of $\uD
3$-$\overline{\uD 3}$ pairs
\begin{equation}
  \mathcal{S}\sim\frac{P\tau_{\uD 3}\kappa_{10}^{2}}
  {\left(g_{s}M\alpha'\right)^{2}\tilde{\mathcal{V}}_{\Omega}},
\end{equation}
where $\tau_{\uD 3}$ is the tension of a $\uD 3$ brane and
$\mathcal{\tilde{V}}_{\Omega}$ is the unwarped volume of the $S^{3}$
at the tip\footnote{This value parametrically differs from the
  analogous parameter in the DKM solution~\cite{DeWolfe:2008zy} by
  $\mathcal{S}\sim\mathcal{S}^{\mathrm{DKM}}\ve^{-8/3}$.}.  Since our
interest will be only in the parametric dependence of the gaugino
mass, we will not need the more detailed expressions for the solution
found in~\cite[Section 3.3]{MSS1}.

The fractional $\uD 3$-branes of the KS solution, together with the
additional $\uD 3$-$\overline{\uD 3}$ pairs, produce non-trivial warping
\begin{equation}
  h\bigl(\tau\bigr)=h_{\KS}+\frac{h_{0}}{\tau},
\end{equation}
where
\begin{equation}
  h_{\KS}\bigl(\tau\bigr)=\bigl(g_{s}M\alpha'\bigr)^{2}
  2^{2/3}\ve^{-8/3}I\bigl(\tau\bigr),\qquad
  I\bigl(\tau\bigr)=\int_{\tau}^{\infty}
  \ud x\,\frac{x\coth x-1}{\sinh^{2}x}
  \bigl(\sinh 2x-2x\bigr)^{1/3},
\end{equation}
and
\begin{equation}
  h_{0}\sim\left(g_{s}M\alpha'\right)^{2}\ve^{-8/3}\mathcal{S}.
\end{equation}
The geometry exhibits a curvature singularity at $\tau=0$ where the
Ricci scalar behaves as
\begin{equation}
  R\sim\frac{\mathcal{S}}{g_{s} M \tau}.
\end{equation}
The lower bound~\eqref{eq:taulowerbound} also follows from demanding that
the Ricci scalar remains small in string units.

The fluxes are
\begin{subequations}
  \begin{align}
    B_{2}=&\frac{g_{s}M\alpha'}{2}\left[ f\left(\tau\right)g_{1}\wedge
      g_{2}+
      k\left(\tau\right)g_{3}\wedge g_{4}\right], \\
    H_{3}=&\frac{g_{s}M\alpha'}{2}\bigl[
    \ud\tau\wedge\bigl(f'\bigl(\tau\bigr)g_{1}\wedge g_{2}+
    k'\bigl(\tau\bigr)g_{3}\wedge g_{4}\bigr)+
    \frac{1}{2}\bigl(k\bigl(\tau\bigr)\!-f\!\bigl(\tau\bigr)\bigr)
    g_{5}\wedge\bigl(g_{1}\wedge g_{3}+g_{2}\wedge g_{4}\bigr)
    \bigr], \\
    F_{3}=&\frac{M\alpha'}{2}\left[ \left(1-F\left(\tau\right)\right)
      g_{5}\wedge g_{3}\wedge g_{4}+ F\left(\tau\right)g_{5}\wedge
      g_{1}\wedge g_{2} + F'\left(\tau\right)\ud\tau\wedge
      \left(g_{1}\wedge g_{3}+g_{2}\wedge g_{4}\right)\right],
  \end{align}
\end{subequations}
with
\begin{equation}
  f\left(\tau\right)=f_{\KS}+f_{0},\quad
  k\left(\tau\right)=k_{\KS}+\frac{k_{0}}{\tau^{2}},\quad
  F\left(\tau\right)=F_{\KS}+\frac{F_{0}}{\tau},
\end{equation}
where the KS solution is
\begin{align}
  f_{\KS}\left(\tau\right)=& \frac{\tau\coth\tau -1}{2\sinh\tau}
  \left(\cosh\tau -1\right),\quad &k_{\KS}\left(\tau\right)=&
  \frac{\tau\coth\tau -1}{2\sinh\tau}
  \left(\cosh\tau +1\right),\notag \\
  F_{\KS}\left(\tau\right)=& \frac{\sinh\tau-\tau}{2\sinh\tau},
\end{align}
and again
\begin{equation}
  f_{0}\sim k_{0}\sim F_{0}\sim\mathcal{S}.
\end{equation}
These source the R-R 5-form,
\begin{subequations}
  \begin{equation}
    F_{5}=\bigl(1+\ast_{10}\bigr)\mathcal{F}_{5},\qquad
    \mathcal{F}_{5}=
    \frac{g_{s}M^{2}\alpha'^{2}}{4}
    \ell\left(\tau\right)
    g_{1}\wedge g_{2}\wedge g_{3}\wedge g_{4}\wedge g_{5},
  \end{equation}
  where
  \begin{equation}
    \ell\bigl(\tau\bigr)= f\left(1-F\right)+kF.
  \end{equation}
\end{subequations}
For the choice of parameters implicitly considered here, this
solution, like that in~\cite{DeWolfe:2008zy}, does not introduce a net
amount of charge localized at the tip since the $\uD 3$ and
$\overline{\uD 3}$-branes are added in pairs.  However, $H_{3}$ and
$F_{3}$ give rise to an effective $\uD 3$ charge which is dual to the
scale dependent effective 't Hooft coupling
\begin{equation}
  g_{s}N_{\mathrm{eff}}\bigl(\tau\bigr)
  \sim \bigl(g_{s}M\alpha'\bigr)^{2}\ell\bigl(\tau\bigr).
\end{equation}

Finally, the SUSY-breaking 3-form fluxes give a non-trivial source for
the dilaton,
\begin{equation}
  \Phi\bigl(\tau\bigr)=\log g_{s}+\Phi_{0}\tau,
\end{equation}
where $\Phi_{0}\sim\mathcal{S}$.  The axion in both KS and this
perturbation is trivial, $C=0$.

The $\mathbb{Z}_{2}$ $R$-symmetry is realized geometrically as a shift
in an angle $\psi\to\psi+2\pi$ (as briefly reviewed in
Appendix~\ref{app:conifold}, $\psi$ ranges from $0$ to $4\pi$).  Since
the $\uD 3$-$\overline{\uD 3}$ pairs are smeared over the angular
directions, the expressions for the bulk fields respect this shift
symmetry.  It is thus reasonable to assume that the SUSY-breaking
state in the dual theory preserves the $\mathbb{Z}_{2}$ $R$-symmetry.

In~\cite{Camara:2004jj,Lust:2008zd,Benini:2009ff}, it was argued that
the existence of a nonvanishing gaugino mass for the world-volume
gauge theory living on a $\uD 7$-brane is related to the existence of
3-form flux with Hodge type $\left(0,3\right)$.  Using the relations
between the 1-form $g_{i}$ and the holomorphic
coordinates~(\ref{eq:deformconifold}) reviewed in
Appendix~\ref{app:conifold}, one can show that indeed the 3-form flux
$G_{3}$ picks up such components in the above perturbation of KS.
Using~\eqref{eq:generalmixed} and~\eqref{eq:generalpure}, the only
non-vanishing component for KS is the $\left(2,1\right)$ component,
\begin{align}
  \label{eq:KS3form}
  G_{3\left(\KS\right)}^{\left(2,1\right)} =\frac{M\alpha'}{2\ve^{6}}
  &\left[\frac{\sinh 2\tau-2\tau}{\sinh^{5}\tau} \left(\bar{z}_{m}\ud
      z_{m}\right)\wedge \left(\epsilon_{ijk\ell}z_{i}\bar{z}_{j}
      \ud z_{k}\wedge\ud\bar{z}_{\ell}\right)\right. \notag \\
  & \left.+\frac{2\left(1-\tau\coth\tau\right)}{\sinh^{4}\tau}
    \left(z_{m}\ud\bar{z}_{m}\right)\wedge
    \left(\epsilon_{ijk\ell}z_{i}\bar{z}_{j} \ud z_{k}\wedge\ud
      z_{\ell}\right)\right].
\end{align}
However, the above perturbation includes non-vanishing values for all
components
\begin{subequations}
  \begin{align}
    \delta G_{3}^{\left(2,1\right)}&
    \sim\frac{\mathcal{S}M\alpha'}{\ve^{6}\tau^{5}}
    \left[c_{1}\left(\bar{z}_{m}\ud z_{m}\right)\wedge
      \left(\epsilon_{ijk\ell}z_{i}\bar{z}_{j} \ud
        z_{k}\wedge\ud\bar{z}_{\ell}\right)
      +c_{2}\left(z_{m}\ud\bar{z}_{m}\right)\wedge
      \left(\epsilon_{ijk\ell}z_{i}\bar{z}_{j}
        \ud z_{k}\wedge\ud z_{\ell}\right)\right], \\
    \delta G_{3}^{\left(1,2\right)}&
    \sim\frac{\mathcal{S}M\alpha'}{\ve^{6}\tau^{5}} \left[c_{3}
      \left(z_{m}\ud \bar{z}_{m}\right)\wedge
      \left(\epsilon_{ijk\ell}z_{i}\bar{z}_{j} \ud
        z_{k}\wedge\ud\bar{z}_{\ell}\right) +c_{4}\left(\bar{z}_{m}\ud
        z_{m}\right)\wedge \left(\epsilon_{ijk\ell}z_{i}\bar{z}_{j}
        \ud
        \bar{z}_{k}\wedge\ud \bar{z}_{\ell}\right)\right], \\
    \delta G_{3}^{\left(3,0\right)}& \sim
    \frac{c_{5}\mathcal{S}M\alpha'}{\ve^{6}\tau^{3}}
    \left(\bar{z}_{m}\ud z_{m}\right)\wedge
    \left(\epsilon_{ijk\ell}z_{i}\bar{z}_{j} \ud z_{k}\wedge\ud
      z_{\ell}\right),
    \label{eq:30component}\\
    \delta G_{3}^{\left(0,3\right)}& \sim
    \frac{c_{6}\mathcal{S}M\alpha'}{\ve^{6}\tau^{3}} \left(z_{m}\ud
      \bar{z}_{m}\right)\wedge
    \left(\epsilon_{ijk\ell}z_{i}\bar{z}_{j} \ud \bar{z}_{k}\wedge\ud
      \bar{z}_{\ell}\right).
  \end{align}
\end{subequations}
where the $c_{i}$ are non-vanishing $\mathcal{O}\left(1\right)$
coefficients whose exact values we will not need.  In contrast, only
$\left(1,2\right)$ and $\left(2,1\right)$ components appeared in the
large radius solution of~\cite{DeWolfe:2008zy}.

It was shown in~\cite{Benini:2009ff} that if the complex structure of
the space changes, then the existence of $g_{zz}$ and
$g_{\bar{z}\bar{z}}$ components of the metric can give rise to
additional contributions to the gaugino mass.  Such components exist
in this perturbation.  Using~\eqref{eq:generalmetric}, the unwarped
metric for the holographic and internal radial directions for the KS
metric is Calabi-Yau
\begin{equation}
  \ud\tilde{s}_{6}^{2}=\bigl(\partial_{i}\partial_{\bar{j}}
  \mathcal{F}\bigr)\ud z_{i}\ud\bar{z}_{j},\qquad
  \mathcal{F}'\left(\ve^{2}\cosh\tau\right)
  =\ve^{-2/3}K\left(\tau\right),
\end{equation}
while the perturbation to the metric is not even Hermitian with
respect to the original complex structure
\begin{align}
  \bigl(\ve^{4}\sinh^{2}\tau\bigr)
  \delta\left(\ud\tilde{s}_{6}^{2}\right) \sim
  &\frac{\mathcal{S}\ve^{4/3}}{\tau}\bigl[
  d_{1}\bigl(\bigl(\bar{z}_{i}\ud z_{i}\bigr)^{2}+
  \bigl(z_{i}\ud\bar{z}_{i}\bigr)^{2}\bigr)+ d_{2}\bigl(\bar{z}_{i}\ud
  z_{i}\bigr)\bigl(z_{i}\ud\bar{z}_{i}\bigr)\bigr]
  \notag \\
  &+ \mathcal{S}\ve^{10/3}\tau\bigl[d_{3} \bigl(\ud z_{i}\ud
  z_{i}+\ud\bar{z}_{i}\ud\bar{z}_{i}\bigr) +d_{4}\ud
  z_{i}\ud\bar{z}_{i}\bigr],
\end{align}
where the $d_{i}$ are another set of $\mathcal{O}\left(1\right)$
coefficients.

For the purposes of calculating the gaugino mass, it is useful to
introduce another set of holomorphic $1$-forms $\ud
Z_{i}$~\eqref{eq:Zcoordinates}.  Using~\eqref{eq:generalmixedZ}
and~\eqref{eq:generalpureZ}, the components of $G_{3}$ can be written
in these coordinates as
\begin{align}
  G_{3\left(\KS\right)}=-\frac{M\alpha'}{16\sinh^{2}\tau}\bigl[&
  4\bigl(\sinh\tau-\tau\cosh\tau\bigr)\ud\bar{Z}_{1}\wedge
  \ud Z_{2}\wedge\ud Z_{3}\bigr. \notag \\
  &+\bigl(\sinh 2\tau-2\tau\bigr)\bigl( \ud Z_{1}\wedge\ud
  Z_{2}\wedge\ud\bar{Z}_{3}- \ud Z_{1}\wedge\ud\bar{Z}_{2}\wedge\ud
  Z_{3}\bigr)\bigr],
\end{align}
while the perturbation to $G_{3}$ has components
\begin{subequations}
  \begin{align}
    \delta
    G_{3}^{\left(2,1\right)}\sim&-\frac{\mathcal{S}M\alpha'}{4\tau^{2}}
    \bigl[c_{1}\bigl(\ud Z_{1}\wedge\ud Z_{2}\wedge\ud\bar{Z}_{3}- \ud
    Z_{1}\wedge\ud\bar{Z}_{2}\wedge\ud Z_{3}\bigr)
    + c_{2}\, \ud\bar{Z}_{1}\wedge\ud Z_{2}\wedge\ud Z_{3}\bigr], \\
    \delta G_{3}^{\left(1,2\right)}\sim&+
    \frac{\mathcal{S}M\alpha'}{4\tau^{2}}\bigr[
    c_{3}\bigl(\ud\bar{Z}_{1}\wedge\ud\bar{Z}_{2}\wedge\ud Z_{3}
    -\ud\bar{Z}_{1}\wedge\ud Z_{2}\wedge\ud\bar{Z}_{3}\bigr)
    +c_{4}\, \ud Z_{1}\wedge\ud\bar{Z}_{2}\wedge\ud\bar{Z}_{3}\bigr], \\
    \delta
    G_{3}^{\left(3,0\right)}\sim&-\left(c_{5}\mathcal{S}M\alpha'\right)
    \ud Z_{1}\wedge\ud Z_{2}\wedge\ud Z_{3}, \\
    \delta
    G_{3}^{\left(0,3\right)}\sim&+\left(c_{6}\mathcal{S}M\alpha'\right)
    \ud \bar{Z}_{1}\wedge\ud\bar{Z}_{2}\wedge\ud\bar{Z}_{3}.
  \end{align}
\end{subequations}
The metric in these coordinates is~\eqref{eq:internalmetricZ}.  In KS,
this becomes
\begin{equation}
  \ud\tilde{s}_{6}^{2}
  =\frac{\ve^{4/3}}{6 K^{2}}\ud Z_{1}\ud\bar{Z}_{2}+
  \frac{\ve^{4/3}K}{2}\sinh^{2}\frac{\tau}{2}\ud Z_{2}\ud\bar{Z}_{2}
  +\frac{\ve^{4/3}K}{2}\cosh^{2}\frac{\tau}{2}\ud Z_{3}\ud\bar{Z}_{3},
\end{equation}
while the perturbation to the metric is
\begin{align}
  \delta\bigl(\ud
  \tilde{s}_{6}^{2}\bigr)\sim&\mathcal{S}\ve^{4/3}\biggl[
  \frac{\hat{d}_{1}}{\tau}\ud Z_{1}\ud\bar{Z}_{1}+ \hat{d}_{2}\tau\,
  \ud Z_{2}\ud\bar{Z}_{2}+
  \frac{\hat{d}_{3}}{\tau}\ud Z_{3}\ud\bar{Z}_{3} \notag \\
  & + \frac{\hat{d_{4}}}{\tau}\bigl(\ud Z_{1}\ud Z_{1}+\ud\bar{Z}_{1}
  \ud\bar{Z}_{1}\bigr) +\hat{d}_{5}\tau\bigl(\ud Z_{2}\ud
  Z_{2}+\ud\bar{Z}_{2}\ud\bar{Z}_{2}\bigr) +
  \frac{\hat{d_{6}}}{\tau}\bigl(\ud Z_{3}\ud Z_{3}+\ud\bar{Z}_{3}
  \ud\bar{Z}_{3}\big)\biggr],
\end{align}
where the $\hat{d}_{i}$ are
$\mathcal{O}\left(1\right)$ coefficients that can be written in terms of
$d_{i}$.

\section{Gaugino masses from holography}\label{sec:gaugino}

Using the above SUSY-breaking gravity solution, we can now proceed to
calculate the mass of a gaugino living on a stack of $K$ probe $\uD
7$-branes in this geometry. In order to neglect the back reaction of
the $\uD 7$-branes, we take $K\ll P\ll M$.  Although it would be
interesting to calculate the back reaction as
in~\cite{Ouyang:2003df,Benini:2007gx,Benini:2007kg}, such a
calculation would lead to a self-energy problem when we try to
calculate the mass of a gaugino living on the $\uD 7$s.  The
calculation here is similar that of~\cite{Benini:2009ff} though
because of the reduced isometry of the geometry (which in the dual
field theory corresponds to reduced $R$-symmetry in the hidden sector)
it leads to a non-vanishing result even to leading order in
$\mathcal{S}$.

The starting point is the Dirac-like action for a $\uD 7$-brane
presented in~\cite{Martucci:2005rb} based largely
on~\cite{Marolf:2003ye,Marolf:2003vf} and reviewed in
Appendix~\ref{app:conv}.  This action is strictly speaking only valid
in the Abelian (i.e. $K=1$) case, but in the supergravity limit, we do
not expect any deviations for the gaugino mass from the Abelian
result\footnote{The fermionic $\uD p$-brane action should also contain
  a Yukawa-like coupling
  $\tr\left\{\lambda,\left[\Phi,\lambda\right]\right\}$ where the
  $\Phi$ are the transverse fluctuations of $\uD 7$-branes.  However,
  such a term only contributes to the gaugino mass at loop level on
  the gravity side which corresponds to a finite 't Hooft coupling
  effect on the field theory side.  Indeed, this term gives a coupling
  between the gaugino and the meson messengers $\Phi_{n}$ and as shown
  in~\cite{Benini:2009ff} gives a contribution suppressed by the `t
  Hooft coupling.}.  The strategy is to find the effective mass for
the gaugino that results from a dimensional reduction of the
world-volume action to $\mathbb{R}^{1,3}$, which in the dual field
theory corresponds to calculating the mass resulting from all planar
diagrams in the 't~Hooft limit.  We begin with an analysis of
contributions to the gaugino mass from $3$-form flux (some of which
are non-vanishing).  Similar considerations were performed
in~\cite{Camara:2004jj} and~\cite{Benini:2009ff}.  However, \textit{a
  priori} there could be additional contributions from other bulk
fields which we consider towards the end of this section.

In the KS background, the 3-form flux is ISD $\left(2,1\right)$ and
the gaugino remains massless~\cite{Camara:2004jj}.  Thus contributions
to the gaugino mass will come from the non-SUSY perturbations to KS.
Since the solution is known only to leading order in $\mathcal{S}$, we
will be interested only in contributions to the gaugino mass that are
also linear in $\mathcal{S}$.

\subsection{Contributions from the 3-form flux}

The fermionic action is written in terms of a bispinor
\begin{equation}
  \Theta = \begin{pmatrix} \theta \\ \tilde{\theta}\end{pmatrix},
\end{equation}
where $\theta$ and $\tilde{\theta}$ are 10D Majorana-Weyl spinors of positive
chirality.  For the probe $\uD 7$-branes, the contribution to the Dirac action
from the $3$-form flux can be written as a trace over gauge indices
\begin{multline}
  \label{eq:3formaction}
  S_{\uD 7}^{\left(3\right)}=\frac{\ui\tau_{\uD 7}g_{s}^{-1/2}}{8}
  \int\ud^{8}\xi\, \ue^{3\Phi/2}
  \sqrt{\abs{\det\mathcal{M}}} \\
  \times\tr\biggl\{\overline{\Theta} P_{-}^{\uD
    7}\biggl[2\mathcal{G}_{3}^{+}
  +\left(\hat{\mathcal{M}}^{-1}\right)^{\alpha\beta}\Gamma_{\beta}
  \left(\mathcal{G}_{3}^{-}\Gamma_{\alpha}
    +\frac{1}{2}\Gamma_{\alpha}\left(\mathcal{G}_{3}^{-}-
      \mathcal{G}_{3}^{+}\right)\right)\biggr]\Theta\biggr\},
\end{multline}
where in the absence of world-volume flux,
\begin{equation}
  \M_{\alpha\beta}=\gamma_{\alpha\beta}+g_{s}^{1/2}\ue^{-\Phi/2}b_{\alpha\beta},
\end{equation}
with $\gamma$ and $b$ the pullbacks of the metric and the NS-NS 2-form
and
\begin{equation}
  \hat{\mathcal{M}}_{\alpha\beta}=\begin{pmatrix}
   \M_{\beta\alpha} & 0\\ 0 & \M_{\alpha\beta}\end{pmatrix},
\end{equation}	
where $\xi^{\alpha}$ are the world-volume coordinates and tensors with
indices $\alpha,\beta$ denote pullbacks onto the world-volumes of the
branes.  $x^{\mu}$ denotes a coordinate in the four large spacetime
dimensions while $x^{a}$ are coordinates on the radial direction and
the internal angular directions.  When acting on the gaugino in the
supersymmetric case, the projection operator can be written as $P^{\uD
  7}_{\pm}=\frac{1}{2}\bigl(1\mp\Gamma_{\uD 7}\bigr)$
with~\cite{Martucci:2005rb,Marchesano:2008rg}
\begin{equation}
  \label{eq:projop}
  \Gamma_{\uD 7}=\begin{pmatrix}
    0 & \ui\Gamma_{\left(8\right)} \\
    -\ui\Gamma_{\left(8\right)} & 0\end{pmatrix},
\end{equation}
where $\Gamma_{\left(8\right)}$ is the usual 8D chirality operator.
The solution presented in Section~\ref{subsec:gravsol} is no longer BPS,
but the deviation from Eq.~(\ref{eq:projop}) essentially gives a
mixing term and so contributes to the gaugino mass at higher order in
$\mathcal{S}$.  Finally the contribution from the 3-form flux is
\begin{equation}
  \mathcal{G}_{3}^{\pm}
  =\frac{1}{3!}\left(\tilde{F}_{MNP}\sigma_{1}
    \pm\ue^{-\Phi}H_{MNP}\sigma_{3}\right)\Gamma^{MNP}.
\end{equation}

As is well known, the fermionic part of the action has a redundant
description of the fermionic degrees of freedom known as
$\kappa$-symmetry.  We choose to eliminate the redundancy by taking the
particular $\kappa$-fixing condition $\tilde{\theta}=0$
\begin{equation}
  \label{eq:kappafix}
  \Theta=\begin{pmatrix} \theta \\ 0\end{pmatrix}.
\end{equation}
To leading order in $\mathcal{S}$, we can take the gaugino
wavefunction to be unperturbed by the addition of the $\overline{\uD
  3}$-branes in which case it is given by~\cite{Marchesano:2008rg}
\begin{equation}
  \label{eq:wf}
  \theta\left(x^{\alpha}\right)=\lambda\left(x^{\mu}\right)\otimes
  h^{3/8}\eta\left(x^{a}\right),
\end{equation}
where $\eta$ is covariantly constant with respect to the underlying
Calabi-Yau (i.e. deformed conifold) metric and is annihilated by the
holomorphic $\Gamma$-matrices $\Gamma_{z}$~\cite{Camara:2003ku}.  Taking
$\theta$ to have negative 6D chirality, the 4D chirality is also negative
\begin{equation}
  \Gamma_{\left(4\right)}\theta=\ui\Gamma^{0123}\theta=-\theta.
\end{equation}
As shown in~\cite{Marchesano:2008rg}, the gaugino has positive
chirality with respect to the chirality operator for the internal
4-cycle wrapped by the $\uD 7$-brane, $\Gamma^{\mathrm{extra}}$.  This
gives a positive 8D chirality.
\begin{equation}
  \Gamma_{\left(8\right)}\theta=-\Gamma_{\left(4\right)}
  \Gamma^{\mathrm{extra}}\theta=\theta.
\end{equation}

Using the above choice of $\kappa$-fixing, we find that the
action~(\ref{eq:3formaction}) is
\begin{align}
  \label{eq:3formactionb}
  S_{\uD 7}^{\left(3\right)} =-&\frac{\tau_{\uD 7}g_{s}^{-1/2}}{8\cdot
    3!}  \int\ud^{8}\xi\, \ue^{3\Phi/2}\sqrt{\abs{\det \M}}
  \tr\biggl\{\bar{\theta}\biggl[G_{MNP}\Gamma^{MNP} \notag \\
  & +\bigl(\M^{-1}\bigr)^{\left(\alpha\beta\right)}\Gamma_{\beta}
  \biggl(G^{\ast}_{MNP}\Gamma^{MNP}\Gamma_{\alpha} +\frac{1}{2}
  \Gamma_{\alpha}\bigl(G^{\ast}_{MNP}-G_{MNP}\bigr)
  \Gamma^{MNP}\biggr)\biggr] \notag \\
  &+ \bigl(\M^{-1}\bigr)^{\left[\alpha\beta\right]}\Gamma_{\beta}
  \biggl(G_{MNP}\Gamma^{MNP}\Gamma_{\alpha}+\frac{1}{2}
  \Gamma_{\alpha}\bigl(G_{MNP}-G^{\ast}_{MNP}\bigr)
  \Gamma^{MNP}\biggr)\biggr]\theta\biggr\},
\end{align}
with
\begin{equation}
  G_{3}=\tilde{F}_{3}-\ui\ue^{-\Phi}H_{3},\qquad
  G_{3}^{\ast}=\tilde{F}+\ui\ue^{-\Phi}H_{3},
\end{equation}
and $\left(\alpha \beta\right)$ and
$\left[\alpha\beta\right]$ indicate symmetrization
and anti-symmetrization over the indices.

To find contributions to the gaugino mass, we must consider
perturbations to the fields in~(\ref{eq:3formactionb}).  The
perturbations to consider are those of the measure
$\ue^{3\Phi/2}\sqrt{\abs{\det \M}}$, the metric $g_{MN}$,
$\M_{\alpha\beta}$, and the $3$-form flux $G_{3}$ (as discussed below,
perturbations of the $\Gamma$-matrices contribute only higher order
terms to the gaugino mass).  Moreover to leading order in
$\mathcal{S}$, we need only to consider the perturbations to one of
these at a time.

\subsubsection{Contributions from the perturbed 3-form flux}
	
We first consider the contributions from the perturbed flux but
unperturbed metric and in particular consider the term
\begin{equation}
  \tr\biggl\{\bar{\theta}G_{MNP}\Gamma^{MNP}\theta\biggr\}.
\end{equation}
$G_{3}$ has legs only on the holographic and internal directions so
\begin{equation}
  \tr\biggl\{\bar{\theta}G_{MNP}\Gamma^{MNP}\theta\biggr\}
  =\tr\bigl(\lambda^{2}\bigr)
  \tilde{g}^{mn}\tilde{g}^{sr}\tilde{g}^{pq}G_{msp}
  \eta^{T}\tilde{\Gamma}_{nrq}\eta,
\end{equation}
where we have used~(\ref{eq:wf}) and related the warped
$\Gamma$-matrices to the unwarped ones
\begin{equation}
  \Gamma_{m}=h^{1/4}\tilde{\Gamma}_{m},
\end{equation}
and $\tilde{g}$ is the unwarped bulk metric.  Since
$\tilde{\Gamma}_{z}\eta=0$, this becomes
\begin{equation}
  \label{eq:gammastoOmega}
  \tr\biggl\{\bar{\theta}G_{MNP}\Gamma^{MNP}\theta\biggr\}
  =\tr\bigl(\lambda^{2}\bigr)
  \tilde{g}^{i\bar{i}'}\tilde{g}^{j\bar{j}'}\tilde{g}^{k\bar{k}'}
  G_{ijk}\eta^{T}\tilde{\Gamma}_{\bar{i}'\bar{j}'\bar{k}'}\eta,
\end{equation}
where $i,j,k$ are holomorphic indices and $\bar{i},\bar{j},\bar{k}$
are anti-holomorphic.  Terms that involve $\Gamma$-matrices of mixed
types (e.g. $\Gamma_{\bar{i}}\Gamma_{j}\Gamma_{\bar{k}}$) give rise to
mixing terms and so contribute to the gaugino mass at higher order in
the perturbation.  Eq. \eqref{eq:gammastoOmega} implies that in addition
to the $\left(0,3\right)$ contribution to the gaugino mass argued to
exist in~\cite{Camara:2004jj} and coming from~\eqref{eq:3formactione},
there is a contribution from the $\left(3,0\right)$-component.  $\eta$
is covariantly constant with respect to the Calabi-Yau metric which
allows us to write\footnote{We note that here it is especially
  important that the derivative appearing in the Dirac-like action is
  the pullback of the covariant derivative built from the bulk metric
  and not the covariant derivative built from the pullback of the
  metric.}
\begin{equation}
  \eta^{T}\tilde{\Gamma}_{\bar{i}\bar{j}\bar{k}}\eta =
  \bar{\Omega}_{\bar{i}\bar{j}\bar{k}},
\end{equation}
where $\Omega$ is the holomorphic 3-form of the underlying Calabi-Yau.
Thus, there is a contribution to the gaugino mass of the form
\begin{equation}
  \label{eq:3formactionc}
  -\frac{\tau_{\uD 7}g_{s}^{-1/2}}{8\cdot 3!}
  \int_{\mathrm{R}^{1,3}}\ud^{4}x\, \tr\left(\lambda^{2}\right)
  \int_{\Sigma_{4}}\ud^{4}x\, \ue^{3\Phi/2}\sqrt{\abs{\det\M}}
  \bar{\Omega}^{\widetilde{ijk}}G_{ijk}.
\end{equation}
where $\tilde{i}$, etc denote indices raised with the unwarped metric
$\tilde{\gamma}$ and $\Sigma_{4}$ denotes the 4-cycle wrapped by the
$\uD 7$.

Since we have only calculated the perturbations due to the
$\overline{\uD 3}$-branes as a small $\tau$ expansion, we cannot
calculate the gaugino mass exactly and will therefore only be
interested in a parametric dependence. Using~(\ref{eq:30component})
and~(\ref{eq:3form}), we find
\begin{equation}
  \label{eq:omegag03}
  \frac{1}{3!}\bar{\Omega}^{\widetilde{ijk}}
  G_{ijk}
  \sim \frac{M\alpha'}{\ve^{2}\tau}\mathcal{S}+
  \mathcal{O}\left(\tau\right).
\end{equation}
The terms that are higher order in $\tau$ have been omitted since the
integral in~\eqref{eq:3formactionc} will receive contributions only
for small $\tau$.

Because we are expanding to linear order in $\mathcal{S}$, all other
fields are set to their background (KS) values.  Expressions for the
pullbacks of the metric and NS-NS $2$-form are given in
Appendix~\ref{app:pullbacks}.  Even though it is possible to write an
exact expression for $\det \M$, it is relatively complex and because
we are interested only in the parametric dependence, we will consider
only the behavior for small $\tau$.  To illustrate the approximation
we use for $\det \M$ and other fields, we first consider the
determinant of the induced metric $\gamma$ which has a simpler exact
expression.  In~\cite{Benini:2009ff} it was shown that the determinant
of the pulled-back metric is
\begin{equation}
  \label{eq:kupersteinmeasure}
  \gamma=\frac{K^{4}\left(\mu^{2}-\ve^{2}\right)^{4}}{16\,\ve^{8/3}}
  K_{2}^{2}\cosh^{2}\frac{\rho}{2}\sinh^{2}\frac{\rho}{2},
\end{equation}
where
\begin{subequations}
  \begin{align}
    \ve^{2}\cosh\tau=&\left(\mu^{2}-\ve^{2}\right)\cosh\rho+\mu^{2}, \\
    K\left(\tau\right)=&\frac{\left(\sinh 2\tau-2\tau\right)^{1/3}}
    {2^{1/3}\sinh\tau}, \\
    K_{2}\left(\tau\right)=&\cosh\rho
    -\frac{\left(\mu^{2}-\ve^{2}\right)\sinh^{2}\rho}{\ve^{2}\sinh^{2}\tau}
    \left(\cosh\tau-\frac{2}{3K^{3}}\right).
  \end{align}
\end{subequations}
The stack of $\uD 7$s extends to a minimum value of $\tau$ given by
\begin{equation}
  \tau_{\mathrm{min}}=2\mathrm{arccosh}\frac{\mu}{\ve},
\end{equation}
and the integral~\eqref{eq:3formactionc} should be dominated by
contributions from $\tau$ near this value since the SUSY-breaking
fluxes are peaked at small $\tau$.
Expanding~(\ref{eq:kupersteinmeasure}) about
$\tau=\tau_{\mathrm{min}}$, we find
\begin{equation}
  \gamma=\frac{1}{64\cdot 2^{1/3}\ve^{2/3}}
  \frac{\left(\mu^{2}-\ve^{2}\cosh\tau_{\mathrm{min}}\right)^{3}}
  {\sinh^{3}\tau_{\mathrm{min}}}
  \left(\sinh 2\tau_{\mathrm{min}}-2\tau_{\mathrm{min}}\right)^{4/3}
  \left(\tau-\tau_{\mathrm{min}}\right)
  +\mathcal{O}\left(\left(\tau-\tau_{\mathrm{min}}\right)^{2}\right).
\end{equation}
To trust that the small $\tau$ expansion is good, we must have that
$\tau_{\mathrm{min}}<1$ which implies that $\mu$ cannot be much larger
than $\ve$.  For small $\tau_{\mathrm{min}}$,
\begin{equation}
  \label{eq:muexpansion}
  \mu\approx\ve\left(1+\frac{1}{8}\tau_{\mathrm{min}}^{2}\right).
\end{equation}
Then $\gamma$ takes the approximate form
\begin{equation}
  \label{eq:detgamma}
  \gamma\approx\frac{2^{5/6}}{3^{4/3}}\ve^{16/3}\tau_{\mathrm{min}}^{7}
  \left(\tau-\tau_{\mathrm{min}}\right).
\end{equation}
where higher order terms in $\tau_{\mathrm{min}}$ have been
dropped.  
Following a similar process for $\mathcal{M}$ gives the same
parametric dependence
\begin{equation}
  \sqrt{\abs{\det{\M}}}\sim \ve^{8/3}\tau_{\mathrm{min}}^{7/2}
    \left(\tau-\tau_{\mathrm{min}}\right)^{1/2}.
\end{equation}
  
Combining this with~(\ref{eq:omegag03}),~(\ref{eq:3formactionc})
becomes
\begin{equation}
  \label{eq:3formactiond}
  \sim-\tau_{\uD 7}g_{s}\int_{\mathbb{R}^{1,3}}
  \ud^{4}x\, \tr\bigl(\lambda^{2}\bigr)
  \int_{\tau_{\mathrm{min}}}^{\tau_{\mathrm{max}}}\ud\tau\, \ve^{8/3}
  \tau_	{\mathrm{min}}^{7/2}
  \left(\tau-\tau_{\mathrm{min}}\right)^{1/2}
  \frac{M\alpha'}{\ve^{2}\tau}\mathcal{S},
\end{equation}
where we have omitted the angular integrals since they do not
contribute to the parametric dependence and where
$\tau_{\mathrm{max}}$ represents some UV cutoff for the field theory.
Defining $t=\tau/\tau_{\mathrm{min}}$, we find
\begin{equation}
  \sim -\tau_{\uD 7}g_{s} M\alpha'\ve^{2/3}\tau_{\mathrm{min}}^{4}\mathcal{S}	\int_{\mathbb{R}^{1,3}}\ud^{4}x\, \tr\bigl(\lambda^{2}\bigr).
\end{equation}

In order to extract the mass, the field needs to be canonically
normalized.  The 4D kinetic term is given by
\begin{equation}
  \ui\tau_{\uD 7}g_{s}^{-1}\int\ud^{8}\xi\, \ue^{\Phi}
  \sqrt{\abs{\det \M}}\bar{\Theta}P_{-}^{\uD 7}
  g^{\mu\nu}\Gamma_{\mu}\partial_{\nu}\Theta
  =\frac{\ui\tau_{\uD 7}}{2}\int_{\mathbb{R}^{1,3}}\ud^{4}x\, 
  \tr\left(\lambda\slashed{\partial}\lambda\right)
  \int_{\Sigma_{4}}\ud^{4}x\, \sqrt{\abs{\det\M}}h.
\end{equation}
The 4D gauge coupling which follows from dimensional reduction of the
bosonic part of the $\uD 7$ action and is identified with the visible
sector $\SU{K}$ coupling is given by\footnote{Note that the value for
  $g_{\vis}$ is sensitive to the perturbation to the geometry, but
  this effect can be neglected to leading order in the perturbation.}
\begin{equation}
  \frac{1}{g_{\vis}^{2}}=\tau_{\uD 7}\left(2\pi\alpha'\right)^{2}
  \int_{\Sigma_{4}}\ud^{4}x\, \sqrt{\abs{\det\M}}h,
\end{equation}
so the kinetic term can be expressed as
\begin{equation}
  \label{eq:kineticterm}
  \frac{1}{8\pi^{2}\alpha'^{2}g_{\vis}^{2}}\int_{\mathbb{R}^{1,3}}\ud^{4}x\,
  \tr\left(\lambda\slashed{\partial}\lambda\right).
\end{equation}
Canonically normalizing the field amounts to dividing by the prefactor
of~\eqref{eq:kineticterm} so~(\ref{eq:3formactiond}) gives a
contribution to the gaugino mass
\begin{equation}
  \delta m_{1/2}\sim g_{s}M\ve^{2/3}\tau_{\mathrm{min}}^{4}
  g_{\vis}^{2}\mathcal{S}.
\end{equation}
In~\cite{DeWolfe:2008zy}, the parameter $\mathcal{S}$ was related to
the vacuum energy in the dual field theory
\begin{equation}
  \mathcal{S}\sim\mathcal{S}^{\mathrm{DKM}}\ve^{-8/3}\sim
  \biggl(\frac{\Lambda_{\s}}{\Lambda_{\ve}}\biggr)^{4},
\end{equation}
so that
\begin{equation}
  \label{eq:gauginomass1}
  \delta m_{1/2}\sim g_{\mathrm{vis}}^{2}\lambda\left(\Lambda_{\ve}\right)
  \frac{\Lambda_{\s}^{4}}{\Lambda_{\ve}^{3}}
  \left(\left(\frac{m_{\chi}}{\Lambda_{\ve}}\right)^{3/2}-1\right)^{2},
\end{equation}
where $\lambda\left(\Lambda_{\ve}\right)= g_{s}M$ is
the 't~Hooft coupling of the hidden sector $\SU{M}$ in the far IR,
$m_{\chi}=\mu^{2/3}$ is the messenger mass and
$\Lambda_{\ve}=\ve^{2/3}$ is the confining scale.

Another contribution resulting from perturbing only the $3$-form flux
potentially comes from
\begin{equation}
  \label{eq:3formactione}
  \tr\biggl\{\bar{\theta}
  \bigl(\M^{-1}\bigr)^{\left(\alpha\beta\right)}\Gamma_{\beta}
  \biggl(G^{\ast}_{MNP}\Gamma^{MNP}\Gamma_{\alpha}
  +\frac{1}{2}
  \Gamma_{\alpha}\bigl(G^{\ast}_{MNP}-G_{MNP}\bigr)
  \Gamma^{MNP}\biggr)\biggr]\theta\biggr\}.
\end{equation}
After some manipulation of the $\Gamma$-matrices, this can be written as
\begin{equation}
  \label{eq:3formactionf}
  -\tr\biggl\{\bar{\theta}\biggl(
  \frac{1}{2}\bigl(\mathcal{M}^{-1}\bigr)^{\left(\alpha\beta\right)}
  \gamma_{\alpha\beta}
  \bigl(G_{MNP}+G_{MNP}^{\ast}\bigr)\Gamma^{MNP}
  -6\bigl(\mathcal{M}^{-1}\bigr)^{\left(\alpha\beta\right)}\Gamma_{\alpha}\Gamma_{MN}
  G_{\beta}^{\ast MN}\biggr)\theta\biggr\}.
\end{equation}
Since $\gamma$ and $\mathcal{M}^{-1}$ are unperturbed, they satisfy
\begin{equation}
  \label{eq:traceM}
  \bigl(\M^{-1}\bigr)^{\left(\alpha\beta\right)}\gamma_{\alpha\beta}
  =8 - \frac{2^{2/3}\tau_{\mathrm{min}}^{2}}{3^{4/3}a_{0}}
  + \mathcal{O}\left(\tau_{\mathrm{min}}^{4}\right),
\end{equation}
where we have used the pullbacks presented in
Appendix~\ref{app:pullbacks}.  Thus to leading order in
$\tau_{\mathrm{min}}$, the first two terms of~\eqref{eq:3formactionf},
which couple to the $\left(3,0\right)$ and $\left(0,3\right)$ parts of
$G_{3}$ respectively, result in contributions to the gaugino mass that
are parametrically the same as~(\ref{eq:gauginomass1}).  The third
term of~\eqref{eq:3formactionf} can be cast as
\begin{equation}
  \label{eq:3formactiong}
  \tr\biggl\{\bar{\theta}\frac{\partial x^{p}}{\partial\xi^{a}}
  \frac{\partial x^{q}}{\partial\xi^{b}}
  \bigl(\M^{-1}\bigr)^{\left(ab\right)}g^{mn}g^{st}G_{pnt}^{\ast}
  \Gamma_{q}\Gamma_{ms}\theta\biggr\},
\end{equation}
where the $\uD 7$-brane world-volumes are specified by
$x^{M}=x^{M}\left(\xi^{\alpha}\right)$.  Using the gaugino
wavefunction and the relation between the warped and unwarped
$\Gamma$-matrices, this becomes (up to combinatorial factors)
\begin{equation}
  \label{eq:3formactionh}
  \tr\bigl(\lambda^{2}\bigr)
  \frac{\partial x^{p}}{\partial\xi^{\alpha}}
  \frac{\partial x^{q}}{\partial\xi^{\beta}}
  \left(\tilde{\M}^{-1}\right)^{\alpha\beta}\tilde{g}^{mn}\tilde{g}^{st}
  G_{pnt}^{\ast}\bar{\Omega}_{qms}.
\end{equation}
where we have defined the ``unwarped'' NS-NS tensor
\begin{equation}
  \tilde{M}_{ab}=\tilde{\gamma}_{ab}+h^{-1/2}g_{s}^{1/2}\ue^{-\Phi/2}b_{ab}.
\end{equation}

Because of the non-trivial embedding, this term is more difficult to
compute.  However, a similar computation was considered
in~\cite{Benini:2009ff} in the KT region where it was useful to
introduce holomorphic 1-forms that are analogous
to~(\ref{eq:Zcoordinates}).  Since the world-volumes of the $\uD
7$-branes are specified by the holomorphic
condition~(\ref{eq:kupersteincondition}), the unperturbed induced
metric is Hermitian.  Similarly, the fact that the $\uD 7$-branes are
supersymmetric in the KS background implies that $b_{2}$ is
$\left(1,1\right)$~\cite{Marino:1999af,Gomis:2005wc} and therefore
$\M$ has non-vanishing values only for the components with one
holomorphic and one anti-holomorphic index.
Then~\eqref{eq:3formactiong} can be written
\begin{equation}
  \label{eq:3formactioni}
  \tr\bigl(\lambda^{2}\bigr)
  \frac{\partial Z^{I}}{\partial\zeta^{\sigma}}
  \frac{\partial \bar{Z}^{\bar{I}'}}{\partial \bar{\zeta}^{\bar{\rho}}}
  \bigl(\tilde{\M}^{-1}\bigr)^{\left(\sigma\bar{\rho}\right)}
  \tilde{g}^{J\bar{J}'}\tilde{g}^{K\bar{K}'}
  G^{\ast}_{IJK}\bar{\Omega}_{\bar{I}'\bar{J}'\bar{K'}},
\end{equation}
where $I,J,K$ indicate the coordinates used in~(\ref{eq:Zcoordinates})
and $\zeta$ are some complex coordinates on $\Sigma_{4}$ whose exact
form we will not need.  To leading order,
\begin{equation}
  \label{eq:Gomegacontraction1}
  \tilde{g}^{J\bar{J}'}\tilde{g}^{K\bar{K}'}
  G^{\ast}_{IJK}\bar{\Omega}_{\bar{I}'\bar{J}'\bar{K'}}
  \sim\frac{M\alpha'\mathcal{S}}{\ve^{2/3}}
  \begin{pmatrix} \tau^{-1} & & \\
    & \tau & \\
    & & \tau^{-1}
  \end{pmatrix},
\end{equation}
where $I=1,2,3$ and where terms higher order in $\tau$ have been
dropped.

To precisely calculate~\eqref{eq:3formactioni}, we would need to
transform from these coordinates to $Z^{I}$, taking into account the
non-trivial pullback.  However, since we are only interested in the
leading parametric dependence, it will suffice to consider the
component of the symmetrized $\mathcal{\M}^{-1}$ which has the leading
$\left(\tau-\tau_{\mathrm{min}}\right)$ and $\tau_{\mathrm{min}}$
behavior.  Using the pullbacks presented in~\ref{app:pullbacks}, this
component is
\begin{equation}
  \label{eq:leadingsympart}
  \bigl(\tilde{\M}^{-1}\bigr)^{h_{2}h_{2}}\sim
  \frac{1}{\tau_{\min}\left(\tau-\tau_{\min}\right)\ve^{4/3}}.
\end{equation}
One can show that in addition to complicated angular dependence, the
coordinate transformation is parametrically effected by multiplication
by $\tau_{\mathrm{min}}^{2}$.  Putting these together, we find that
the leading order behavior is
\begin{equation}
  \label{eq:3formactionj}
  \frac{\partial x^{p}}{\partial\xi^{\alpha}}
  \frac{\partial x^{q}}{\partial\xi^{\beta}}
  \bigl(\tilde{\M}^{-1}\bigr)^{\left(\alpha\beta\right)}
  \tilde{g}^{mn}\tilde{g}^{st}G_{pnt}^{\ast}
  \bar{\Omega}_{qms}\sim
  \frac{M\alpha'}{\ve^{2}\left(\tau-\tau_{\mathrm{min}}\right)}.
\end{equation}
Comparing this to~(\ref{eq:omegag03}) which results
in~(\ref{eq:gauginomass1}), we find that~(\ref{eq:3formactionj})
contributes to the gaugino mass an amount that is parametrically the
same as~\eqref{eq:gauginomass1}.

Further contributions due to perturbed $3$-form flux possibly come from
\begin{equation}
  \label{eq:3formantia}
  \tr\biggl\{\bar{\theta}
  \bigl(\M^{-1}\bigr)^{\left[\alpha\beta\right]}\Gamma_{\beta}
  \biggl(G_{MNP}\Gamma^{MNP}\Gamma_{\alpha}
  +\frac{1}{2}
  \Gamma_{\alpha}\bigl(G_{MNP}-G_{MNP}^{\ast}\bigr)
  \Gamma^{MNP}\biggr)\theta\biggr\}.
\end{equation}
Following similar steps that lead to~\eqref{eq:3formactionf}, this becomes
\begin{equation}
  \label{eq:3formantib}
  -\tr\biggl\{\bar{\theta}\bigl(\M^{-1}\bigr)^{\left[\alpha\beta\right]}
  \Gamma_{\beta}\biggl(
  \Gamma_{\alpha}\bigl(G_{MNP}+G^{\ast}_{MNP}\bigr)\Gamma^{MNP}
  -6G_{\alpha NP}\Gamma^{NP}\biggr)\theta
  \biggr\}.
\end{equation}
Using the results in~\cite{Marchesano:2008rg}, when acting on the gaugino
\begin{equation}
  \bigl(\M^{-1}\bigr)^{ab}\Gamma_{a}\Gamma_{b}\theta=
  \bigl(\M^{-1}\bigr)^{ba}\Gamma_{a}\Gamma_{b}\theta.
\end{equation}
Since $\left(M^{-1}\right)^{\left[\mu\nu\right]}=0$, this implies
(when acting on the gaugino)
\begin{equation}
  \bigl(\M^{-1}\bigr)^{\left[\alpha\beta\right]}\Gamma_{\alpha}\Gamma_{\beta} 
  \theta=0.
\end{equation}
Thus~\eqref{eq:3formantia} becomes
\begin{equation}
  \label{eq:3formantic}
  \tr\biggl\{\bar{\theta}
  \bigl(\M^{-1}\bigr)^{\left[\alpha\beta\right]}\Gamma_{\beta}
  G_{\alpha NP}\Gamma^{NP}\theta\biggr\}.
\end{equation}
This term involves a contraction similar to~\eqref{eq:Zcoordinates}
\begin{equation}
  \tilde{g}^{J\bar{J}'}\tilde{g}^{K\bar{K}'}
  G_{IJK}\bar{\Omega}_{\bar{I}'\bar{J}'\bar{K}'}
  \sim
  \frac{M\alpha'\mathcal{S}}{\ve^{2/3}}\begin{pmatrix}
    \tau^{-1} & & \\
    & \tau & \\
    & & \tau^{-1}
  \end{pmatrix}.
\end{equation}
The remaining indices are contracted with the anti-symmetric part of
$\M^{-1}$.  For the KS background, the most leading part is (using the
pullbacks in Appendix~\ref{app:pullbacks})
\begin{equation}
  \bigl(\M^{-1}\bigr)^{\left[\rho h_{2}\right]}
  \sim \frac{1}{\ve^{4/3}\tau_{\mathrm{min}}^{1/2}
    \left(\tau-\tau_{\min}\right)^{1/2}}.
\end{equation}
Comparing to~\eqref{eq:leadingsympart}, this is subleading in
$\left(\tau-\tau_{\min}\right)$ and $\tau_{\min}$ and thus will give
a subleading contribution to the gaugino mass.

\subsubsection{Contributions from the perturbed metric}
\label{subsubsec:pertmetric}

We next need to take into account terms that result from perturbing
the metric while leaving the flux unperturbed.  The perturbations to
the metric that are of the form $\delta g_{z\bar{z}}$ or $\delta
g_{\bar{z}z}$ will not contribute to the gaugino mass since, as shown
above, when the metric is Hermitian, the unperturbed
$\left(2,1\right)$-component of $G_{3}$ does not contribute.  However,
as pointed out in~\cite{Benini:2009ff}, when the perturbed metric is
no longer Hermitian with respect to the original complex structure,
there are in general contributions to the gaugino mass from the
components of the 3-form flux with mixed holomorphic and
anti-holomorphic indices.

We again consider the first term in~(\ref{eq:3formaction}),
\begin{equation}
  \label{eq:3formactionk}
  \tr\biggl\{\bar{\theta}G_{MNP}\Gamma^{MNP}\theta\biggr\}
  =\tr\bigl(\lambda^{2}\bigr)
  \tilde{g}^{mn}\tilde{g}^{sr}\tilde{g}^{pq}G_{mnp}
  \eta^{T}\tilde{\Gamma}_{nrq}\eta.
\end{equation}
If the metric is no longer Hermitian, then there is a contribution of
the form
\begin{equation}
  \tr\bigl(\lambda^{2}\bigr)\tilde{g}^{\bar{i}\bar{i'}}
  \tilde{g}^{j\bar{j}'}\tilde{g}^{k\bar{k}'}
  G_{\bar{i}jk}
  \bar{\Omega}_{\bar{i}'\bar{j}'\bar{k}'}.
\end{equation}
We could of course consider the contractions with even more
non-Hermitian parts (i.e. terms with
$\tilde{g}^{\bar{i}\bar{i'}}\tilde{g}^{\bar{j}\bar{j}'}\tilde{g}^{k\bar{k}'}$)
but since the non-perturbed metric is Calabi-Yau, these are higher
order in $\mathcal{S}$.  Using the solution in
Section~\ref{subsec:gravsol}, we find
\begin{equation}
  \tilde{g}^{\bar{i}\bar{i}'}
  \tilde{g}^{\bar{j}j'}\tilde{g}^{\bar{k}k'}
  G_{\bar{i}'j'k'}\bar{\Omega}_{\bar{i}\bar{j}\bar{k}}
  \sim\frac{M\alpha'\mathcal{S}}{\ve^{2}\tau}.
\end{equation}
Comparing to~(\ref{eq:omegag03}), we see that this term contributes an
amount that is parametrically the same as~(\ref{eq:gauginomass1}).

Since we are for now neglecting the change in $\M^{-1}$, the next
group of terms~\eqref{eq:3formactione} can again be written
as~\eqref{eq:3formactionf} but now considering the $3$-form flux to be
unperturbed and the non-Hermitian perturbations to the bulk metric.
The first term in~(\ref{eq:3formactionf}) again contributes
parametrically the same as~(\ref{eq:3formactionk}) since making use
of~\eqref{eq:traceM} it is of a
closely related form.  The third term
of~\eqref{eq:3formactionf} potentially has a contribution from the
non-Hermitian perturbations of $g$,
\begin{equation}
  \label{eq:3formactionn}
  \tr\bigl(\lambda^{2}\bigr)
  \frac{\partial Z^{I}}{\partial\zeta^{\sigma}}
  \frac{\partial \bar{Z}^{\bar{I}'}}{\partial \bar{\zeta}^{\bar{\rho}}}
  \bigl(\M^{-1}\bigr)^{\sigma\bar{\rho}}\tilde{g}^{\bar{J}\bar{J}'}
  \tilde{g}^{K\bar{K}'}
  G^{\ast}_{I\bar{J}K}\bar{\Omega}_{\bar{I}'\bar{J}'\bar{K'}},
\end{equation}
where $\M^{-1}$ is unperturbed.  Writing
$G^{\ast}_{I\bar{J}K}=\bigl(G_{\bar{I}J\bar{K}}\bigr)^{\ast}$, we see
that this is a coupling to the $\left(1,2\right)$ component of
$G_{3}$.  The unperturbed flux for the KS solution is purely ISD
$\left(2,1\right)$, so this term vanishes and does not contribute to
the gaugino mass.  This argument which also applies to the second term
of~\eqref{eq:3formactionf} $G^{\ast}_{MNP}$ when the metric is
perturbed but the flux is not.

Next we consider contributions resulting from the perturbation of the
symmetric part of $\M^{-1}$ in~\eqref{eq:3formactionf}.  Again, only
the perturbations to the purely holomorphic and purely
anti-holomorphic parts could possibly contribute to a gaugino mass
(perturbations to the components of mixed type, for example
$\delta\mathcal{M}_{z\bar{z}}$ do not contribute to the gaugino mass
to leading order in $\mathcal{S}$).  The first two terms couple to the
$\left(3,0\right)$ parts of $G_{3}$ and $G^{\ast}_{3}$ and the third
term couples to the $\left(1,2\right)$ part of $G_{3}$. Since the
unperturbed flux $G_{3}$ $\left(2,1\right)$, all of these
contributions vanish to leading order in $\mathcal{S}$.

There could additionally be contributions from the purely holomorphic
and purely anti-holomorphic perturbations to the anti-symmetric part
of $\M^{-1}$ in~\eqref{eq:3formantib}.  The first two terms
of~\eqref{eq:3formactionf} give a coupling to the $\left(0,3\right)$
and $\left(3,0\right)$ parts of $G_{3}$ which vanish.  However, the
third term gives a coupling to the $\left(2,1\right)$-component of
$G_{3}$ which is non-vanishing in KS.  To leading order in $\tau$, the
contraction of the unperturbed fields gives
\begin{equation}
  \tilde{g}^{J\bar{J}'}\tilde{g}^{K\bar{K}'}
  G_{\bar{I}JK}\bar{\Omega}_{\bar{I}'\bar{J}'\bar{K}'}\sim
  \frac{M\alpha'}{\ve^{2/3}}
  \begin{pmatrix}
    1 & & \\
    & \tau^{2} & \\
    & & 1
  \end{pmatrix}.
\end{equation}
The remaining indices are again contracted with the anti-symmetric
part of $\M^{-1}$.  Since we are interested in only the parametric
dependence, we consider the component of the anti-symmetrized
$\M^{-1}$ with the most singular dependence in $\tau_{\mathrm{min}}$
and $\left(\tau-\tau_{\mathrm{min}}\right)$, focusing on the part
proportional to $\mathcal{S}$ (since the parts not proportional to
$\mathcal{S}$ cannot contribute here).  The leading component is (see
Appendix~\ref{app:pullbacks})
\begin{equation}
  \bigl(\tilde{M}^{-1}\bigr)^{\left[\rho h_{2}\right]}\sim
  \frac{\mathcal{S}}{\tau_{\mathrm{min}}^{7/2}
    \left(\tau-\tau_{\mathrm{min}}\right)^{1/2}\ve^{4/3}}
\end{equation}
Taking into account the coordinate transformations, to leading order
\begin{equation}
  \sqrt{\abs{\det\M}}\frac{\partial x^{p}}{\partial\xi^{\alpha}}
  \frac{\partial x^{q}}{\partial\xi^{\beta}}
  \bigl(\tilde{\M}^{-1}\bigr)^{\left[\alpha\beta\right]}
  \tilde{g}^{mn}\tilde{g}^{st}G_{pnt}
  \bar{\Omega}_{qms}\sim
  M\alpha'\ve^{2/3}\tau_{\mathrm{min}}^{2}\mathcal{S}.
\end{equation}
Comparing to~\eqref{eq:3formactiond} which
yielded~\eqref{eq:gauginomass1}, we get the contribution to the gaugino mass
\begin{equation}
  \delta m_{1/2}\sim g_{s}M\ve^{2/3}\tau_{\mathrm{min}}^{3}
  g_{\mathrm{vis}}^{2}\mathcal{S}.
\end{equation}
In terms of the parameters of the dual field theory,
\begin{equation}
  \label{eq:gauginomass2}
  \delta m_{1/2}\sim g_{\mathrm{vis}}^{2}\lambda\left(\Lambda_{\ve}\right)
  \frac{\Lambda_{\s}^{4}}{\Lambda_{\ve}^{3}}
  \left(\left(\frac{m_{\chi}}{\Lambda_{\ve}}\right)^{3/2}-1\right)^{3/2},
\end{equation}
which parametrically contributes more significantly than the previous
contributions.

\subsubsection{Other 3-form contributions}

For each of the above terms, we have neglected the fact that the
measure $\ue^{3\Phi/2}\sqrt{\abs{\det \M}}$ and the $\Gamma$-matrices
should also be modified in the new geometry.  However, since we are
working to first order in $\mathcal{S}$, perturbing the measure means
to consider the Dirac-like operator (i.e. $\bar{\theta}\cdots\theta$)
to be unperturbed.  Since the unperturbed operator does not give a
mass to the gaugino, perturbing the measure will not contribute to
$m_{1/2}$ to linear order in $\mathcal{S}$.

Perturbations to the holomorphic $\Gamma$-matrices are of the form
\begin{equation}
  \delta\Gamma_{i}\sim a^{j}_{\ i}\Gamma_{j}+
  b^{\bar{j}}_{\ i}\Gamma_{\bar{j}}
\end{equation}
The perturbations proportional to the holomorphic $\Gamma$-matrices
will not give any new contribution to the gaugino mass since the
holomorphic $\Gamma$-matrices annihilate the gaugino.  However, the
terms proportional to the anti-holomorphic $\Gamma$-matrices will
contribute through terms such as
\begin{equation}
  \bar{\theta}g^{i\bar{i}'}g^{j\bar{j}'}g^{k\bar{k}'}
  G_{\bar{i}'jk}\delta\Gamma_{i}\Gamma_{\bar{j}'}\Gamma_{\bar{k}'}\theta
  \sim b^{\ \bar{i}}_{\ell}\overline{\Omega}^{\widetilde{\ell jk}}
  G_{\bar{i}jk}
\end{equation}
where $b^{\ \bar{i}}_{j}=g^{k\bar{i}}g_{j\bar{\ell}}b_{\
  k}^{\bar{\ell}}$.  This term couples to the non-vanishing
$\left(2,1\right)$ component of the KS 3-form flux and is expected to
parametrically contribute the same amount as previous contributions.
Similar arguments apply when considering perturbations to the
anti-holomorphic $\Gamma$-matrices.

In addition to the above effects, one must take into account the fact
that the $\overline{\uD 3}$-branes will interact with the $\uD
7$-branes, though the consideration is very closely related to the above
discussions.  That is, $z_{4}=\mu$ is a volume minimizing condition in
the KS geometry, but when we perturb the geometry, this condition will
no longer hold.  The world-volumes will be slightly perturbed so that
the embedding is specified by
\begin{equation}
  \label{eq:newembedding}
  \mathcal{F}\left(z_{i},\bar{z}_{i};\mu,\ve,\mathcal{S}\right)=0,
\end{equation}
for some function $\mathcal{F}$.  To leading order in perturbation
theory, we can consider a stack of $\uD 7$-branes
satisfying~\eqref{eq:newembedding} in the original KS geometry.  With
this new condition, in general the $\uD 7$-branes will no longer have
a complex structure that is compatible with that of the bulk
geometry\footnote{Indeed, the new world-volumes may not, in general,
  even admit a complex structure.  However, relaxing the assumption
  that the world-volumes admits a complex structure will not effect
  the conclusion of this discussion.}.  For example, the pullback of a
$\left(1,0\right)$-form will not in general be a
$\left(1,0\right)$-form with respect to any world-volume complex
structure.  Since the existence of the gaugino mass depends on the
Hodge types of the fluxes, this might result in a non-vanishing mass
for the gaugino (which is simply the statement that if the $\uD
7$-brane is not holomorphically embedded into the geometry, then it is
not supersymmetric).  However, the relative change in Hodge type is
$\mathcal{O}\left(\mathcal{S}\right)$.  That is, if $w_{\sigma}$ are
complex coordinates on the $\uD 7$ world-volumes, then
\begin{equation}
  \mathrm{P}\left[\ud z\right]\sim \ud w + \mathcal{S}\ud\bar{w}.
\end{equation}
A possible gaugino mass could arise from the term\footnote{To this
  order in perturbation theory, the gaugino wavefunction is
  unperturbed so it is annihilated by $\Gamma$-matrices that are
  holomorphic with respect to the KS complex structure.}
\begin{equation}
  \bigl(\M^{-1}\bigr)^{\left(ab\right)}g^{mn}g^{st}\bar{\theta}
  \Gamma_{a ms}G^{\ast}_{bnt}\theta
  \sim \bigl(\M^{-1}\bigr)^{\left(ab\right)}g^{mn}g^{st}\bar{\Omega}_{ams}
  G^{\ast}_{bnt}.
\end{equation}
Since the complex structure of the world-volumes may be different than
that of the bulk, this is generally of the form
\begin{equation}
  \bigl(\M^{-1}\bigr)^{\left(\sigma\bar{\rho}\right)}g^{i\bar{i}'}g^{j\bar{j}'}\bigl(
  \bar{\Omega}_{\sigma\bar{i}'\bar{j}'}G^{\ast}_{\bar{\rho}ij}
  +\bar{\Omega}_{\bar{\rho}\bar{i}'\bar{j}'}G^{\ast}_{\sigma ij}\bigr),
\end{equation}
where we have chosen the world-volume complex structure such that the
$\M$ is non-vanishing only for components with one holomorphic and one
anti-holomorphic index.  In the KS background however, both of these
terms vanish.  Considering the first term, since
$\bar{\Omega}_{\sigma\bar{i}'\bar{j}'}$ has a change in the complex
structure ($\sigma$ is a holomorphic world-volume index but
$\bar{\Omega}$ is $\left(0,3\right)$ in the bulk), it is proportional
to $\mathcal{S}$.  Therefore, the only part of
$G^{\ast}_{\bar{\rho}ij}$ that contributes is such that the Hodge type
is compatible with the bulk complex structure; that is, the part that
contributes is the part of $G_{3}^{\ast}$ that is $\left(2,1\right)$
with respect to the bulk complex structure as well.  Since the
$\left(2,1\right)$ part of $G_{3}^{\ast}$ is essentially the
$\left(1,2\right)$ part of $G_{3}$ and to this order in perturbation
the flux is unperturbed, $G^{\ast}_{\bar{\rho}ij}=0$.  Similarly, the
second term couples to the $\left(3,0\right)$ and $\left(2,1\right)$
parts of $G^{\ast}_{3}$ (with respect to the bulk complex structure).
Both of these vanish in KS, and so the second term vanishes as well.

The arguments for the vanishing of these terms were very similar to
those for the perturbations to the symmetrized part of $\M$ considered
in Section~\ref{subsubsec:pertmetric}.  An analogous
argument for the anti-symmetric part of $\M^{-1}$ would show that
there is a coupling to the bulk $\left(2,1\right)$ part of $G_{3}$
when the world-volumes are perturbed.  Although it would be necessary to
calculate $\mathcal{F}$ appearing in~\eqref{eq:newembedding} to calculate
this exactly, we expect that it should be parametrically similar
to~\eqref{eq:gauginomass2}.

\subsection{Contributions from the 5-form flux}
\label{sec:5form}

All of the above subsections focused on the contributions related to the
$3$-form flux and were similar to discussions in~\cite{Camara:2004jj}
and~\cite{Benini:2009ff}.  However, in principle there could be additional
contributions from other SUSY-breaking bulk fields.

For example, there is the possibility of a mass arising from the
$5$-form flux.  In the SUSY case, the $5$-form flux is related to the
warp factor so we must consider the spin connection as well.  The
action contains,
\begin{equation}
  \ui\tau_{\uD 7}g_{s}^{-1}\int\ud^{8}\xi\, \ue^{\Phi}
  \sqrt{\abs{\det \M}}\tr\biggl\{
  \bar{\Theta}P_{-}^{\uD 7}\bigl(\hat{\M}^{-1}\bigr)^{\alpha\beta}
  \Gamma_{\beta}\biggl(\nabla_{\alpha}
  + \frac{g_{s}}{16\cdot 5!}\tilde{F}_{NPQRT}\Gamma^{NPQRT}
  \Gamma_{\alpha}\bigl(\ui\sigma_{2}\bigr)\biggr)\Theta\biggr\}.
\end{equation}
$\nabla_{\alpha}$ is the pullback of the covariant derivative which has
components
\begin{align}
  \nabla_{\mu}=&\partial_{\mu}-\frac{1}{8}\Gamma_{\mu}
  \slashed{\partial}\log h, \\
  \nabla_{m}=&\tilde{\nabla}_{m}
  +\frac{1}{8}\Gamma_{m}\slashed{\partial}\log h
  -\frac{1}{8}\partial_{m}\log h,
\end{align}
where $\tilde{\nabla}$ is the covariant derivative with respect to the
unwarped 6D metric which in this subsection we take to be unperturbed.
Following~\cite{Marchesano:2008rg}, and using the $\kappa$-fixing
condition~\eqref{eq:kappafix}, this becomes (when acting on the
gaugino)
\begin{align}
  \label{eq:5formactionb}
  \frac{\ui\tau_{\uD 7}}{2 g_{s}}\int\ud^{8}\xi&\,\ue^{\Phi}
  \sqrt{\abs{\det \M}}
  \tr\biggl\{\bar{\theta}\biggl[\bigl(\M^{-1}\bigr)^{ab}\Gamma_{a}
  \bigl(\tilde{\nabla}_{b}-\frac{1}{8}\partial_{b}\log h\bigr) \notag \\
  &+\frac{\left(g_{s}M\alpha'\right)^{2}}{16}\ell\left(\tau\right)
  \frac{\sqrt{p}}{\sqrt{u}sqh}
  \bigl(\M^{-1}\bigr)^{ab}\bigl(\partial_{a}\tau\bigr)\Gamma_{b}
  -\frac{1}{2}\bigl(1-\frac{1}{4}\bigl(\M^{-1}\bigr)^{ab}
  \Gamma_{a}\Gamma_{b}\bigr)\slashed{\partial}\log h \notag \\
  & +\frac{\left(g_{s}M\alpha'\right)^{2}}{8}\ell\bigl(\tau\bigr)
  \frac{\sqrt{p}}{\sqrt{u}sqh}
  \bigl(1-\frac{1}{4}\bigl(\M^{-1}\bigr)^{ab}\Gamma_{b}\Gamma_{a}\bigr)
  \Gamma^{\tau}\biggr]\theta\biggr\},
\end{align}
where we have omitted the 4D kinetic term since it does not contribute
to a mass term.  Since $\theta$ is a Majorana-Weyl spinor, any
bilinear $\bar{\theta}\Gamma_{M}\theta$ vanishes.  Therefore, since
the gaugino wavefunction behaves as $h^{3/8}\eta$ where $\eta$ is
covariantly constant with respect to
$\tilde{\nabla}$,~\eqref{eq:5formactionb} becomes
\begin{equation}
  \label{eq:5formactionc}
  \frac{\ui\tau_{\uD 7}}{16g_{s}}\int\ud^{8}\xi\, \ue^{\Phi}
  \sqrt{\abs{\det \M}}\tr\biggl\{\bar{\theta}
  \bigl(\M^{-1}\bigr)^{ab}
  \biggl[\Gamma_{a}\Gamma_{b}\slashed{\partial}\log h
  -\frac{\left(g_{s}M\alpha'\right)^{2}}{4}\ell\left(\tau\right)
  \frac{\sqrt{p}}{\sqrt{u}sqh}\Gamma_{b}\Gamma_{a}\Gamma^{\tau}
  \biggr]\theta\biggr\}.
\end{equation}
Following similar arguments for the $3$-form flux above, to obtain the
contribution to the gaugino mass to linear order in $\mathcal{S}$, we
consider perturbations to one field at a time.  Perturbations to $h$,
$\ell$, or any of the metric functions $p$, $u$, or $s$
do not give a contribution as the unperturbed $\M^{-1}$ has
non-vanishing elements only for components with one holomorphic and
one anti-holomorphic index so that
\begin{equation}
  \bar{\theta}\bigl(\M^{-1}\bigr)^{ab}\Gamma_{a}\Gamma_{b}\Gamma^{\tau}\theta,
\end{equation}
consists of mixed holomorphic and anti-holomorphic $\Gamma$-matrices.
A similar argument applies if we consider perturbing the measure
$\ue^{\Phi}\sqrt{\abs{\det\M}}$.

The next potential contribution is from considering the perturbation to
$\M^{-1}$.  Since the perturbed $\mathcal{M}^{-1}$ contains pieces that
are non-Hermitian, this may \textit{a priori} contribute to the gaugino mass.
The remaining fields are not perturbed from their KS values which satisfy
\begin{equation}
  h'\left(\tau\right)=-\frac{\left(g_{s}M\alpha'\right)^{2}}{4}
  \ell\frac{\sqrt{p}}{\sqrt{u}sq},
\end{equation}
so that~\eqref{eq:5formactionc} gives
\begin{equation}
  \frac{\ui\tau_{\uD 7}}{16}
  \int\ud^{8}\xi\, \sqrt{\abs{\det\M}}
  \tr\biggl\{\bar{\theta}
    \bigl(\M^{-1}\bigr)^{ab}\bigl\{\Gamma_{a},\Gamma_{b}\bigr\}
    \slashed{\partial}\log h\, \theta\biggr\}.
\end{equation}
Using the Clifford algebra the term in the trace is
\begin{equation}
  \label{eq:5formMperturbterm}
  \bigl(\M^{-1}\bigr)^{ab}\gamma_{ab}\, \bar{\theta}\slashed{\partial}\log h\, 
  \theta=0,
\end{equation}
where we have again used the fact that $\bar{\theta}\Gamma_{M}\theta=0$.
Perturbations to the $\Gamma$-matrices will give the same
form~\eqref{eq:5formMperturbterm} except $\gamma$, rather than $\M^{-1}$ 
is perturbed, and so the term will also vanish.

Similar to the consideration of the $3$-form fluxes, we must also
consider the effect of the deformation of the world-volumes.  However,
for the $3$-form fluxes the important aspect was the change in
complex structure as a result of the pullback.  In this case, the
complex structure of the pullbacks of the $5$-form flux and the spin
connection are not important for arguing for the vanishing of the
mass.  Therefore, considering the effect of the perturbation of the
world-volumes is equivalent to considering perturbations to the fields
and all of these contributions vanish.

\subsection{Contributions from the perturbed spin connection}

Additional contributions could potentially arise from the perturbed
spin connection.  The 6D manifold is perturbed from the deformed
conifold geometry so that it is no longer conformally Calabi-Yau.
Contained within the $\uD 7$ Dirac-like action is the term
\begin{equation}
  \frac{\ui\tau_{\uD 7}}{2g_{s}}\int\ud^{8}\xi\, 
  \tr\bigl\{\bar{\theta}\bigl(\mathcal{M}^{-1}\bigr)^{ab}
  \Gamma_{a}\tilde{\nabla}_{b}\theta\bigr\}
\end{equation}
where $\tilde{\nabla}$ is the pullback of the covariant derivative
with respect to the unwarped 6D metric.
When $\tilde{\nabla}$ is unperturbed, the fact that $\eta$ is covariantly
constant causes this term to automatically vanish.  Therefore, in considering
non-vanishing contributions, we need only consider perturbations to
$\tilde{\nabla}$.  $\tilde{\nabla}$ is given by
\begin{equation}
  \tilde{\nabla}_{a}=\partial_{a}+\frac{1}{4}\tilde{\omega}_{a}^{\underline{MN}}
  \Gamma_{\underline{M}\underline{N}}.
\end{equation}
where $\tilde{\omega}$ is the spin connection built from the unwarped
6D metric.  Perturbations to the spin connection then give the
contribution
\begin{equation}
  \bigl(M^{-1}\bigr)^{ab}g^{mn}g^{st}
  \delta\tilde{\omega}_{ams}\bar{\Omega}_{bnt},
\end{equation}
where we only consider the perturbations to $\omega$, the unperturbed
part being cancelled by the derivative $\partial_{a}$.  A detailed
calculation shows that this contraction vanishes for the
isometry-preserving perturbation considered here.  Some terms in the
perturbed spin connection have the wrong Hodge type to contract with
$\bar{\Omega}$.  The sum over the remaining terms vanish based on the
symmetries of $\delta\tilde{\omega}$. Thus there are no contributions
to the visible sector gaugino mass coming from considering the
perturbations to the 6D unwarped spin connection.

The components $\nabla_{\mu}$ are also perturbed in this geometry.
However, since this geometry is unperturbed from Minkowski space, the
only perturbation to the covariant derivative comes from the
perturbations to the warp factor which were considered in
Section~\ref{sec:5form}.

\section{Conclusion}\label{sec:discussion}

In this paper, we have used to the language of the gauge-gravity
correspondence to consider the effects the effects of strong coupling
dynamics on a relative of semi-direct gauge mediation.  In particular,
we examined the holographic gauge mediation scenario
of~\cite{Benini:2009ff} where the hidden sector is a cascading
$\SU{N+M}\times \SU{N}$ gauge theory, but considered the regime where
the messenger mass $m_{\chi}$ was comparable (and in fact very close
to) the confinement scale $\Lambda_{\ve}$.  In the gravity dual, this
required use of one of the solutions presented in~\cite{MSS1} which
described the influence of an $\overline{\uD 3}$ on the near-tip
geometry of the warped deformed conifold.  The confining dynamics of
the strongly coupled gauge theory breaks the $R$-symmetry to
$\mathbb{Z}_{2}$, which allows the gaugino to get a mass from physics
above $m_{\chi}$.  To leading order in the SUSY-breaking order parameter
\begin{equation}
  \label{eq:gauginomass}
  \delta m_{1/2}\sim g_{\mathrm{vis}}^{2}
  \lambda\left(\Lambda_{\ve}\right)
  \frac{\Lambda_{\s}^{4}}{\Lambda_{\ve}^{3}}
  \left(\left(\frac{m_{\chi}}{\Lambda_{\ve}}\right)^{3/2}-1\right)^{3/2},
\end{equation}
where $\Lambda_{\ve}$ is the hidden sector confining scale,
$\Lambda_{\s}$ is the vacuum energy, $m_{\chi}$ is the messenger mass,
$g_{\mathrm{vis}}$ is the visible sector gauge coupling, and
$\lambda\left(\Lambda_{\ve}\right)$ is the hidden sector 't~Hooft
coupling evaluated at the scale $\Lambda_{\ve}$ (where the cascade has
ended so the gauge group is the simple group $\SU{M}$).  From the many
possible terms that a priori could have given rise to a non-vanishing
contributions, the only non-vanishing contributions come from the
$3$-form flux on the gravity side of the duality.

There are additional contributions from physics below the messenger
mass.  In particular, $\chi$ and $\tilde{\chi}$ bind into weakly
interacting mesons and the spectrum contains mesons with masses below
$m_{\chi}$.  For $m_{\chi}\gg\Lambda_{\ve}$, this contribution was
calculated in~\cite{Benini:2009ff} resulting in~\eqref{eq:beninimass}.
Although we did not calculate the contribution from the mesons in this
geometry, the mesons are weakly coupled in the large 't Hooft coupling
limit while $\chi$ and $\tilde{\chi}$ are strongly coupled.  Thus we
expect the contribution to the gaugino mass from any one meson to be
highly suppressed by 't~Hooft coupling compared to the contributions
from $\chi$.

The 't~Hooft enhancement of~\eqref{eq:gauginomass} is quite different
than the leading order contribution in the regime $m_{\chi}\gg
\Lambda_{\ve}$.  In this regime, considered in~\cite{Benini:2009ff},
the large $R$-symmetry at high energies suppresses contributions to
the gaugino mass from physics above the messenger mass $m_{\chi}$ and
the leading order contribution comes from the 't~Hooft suppressed
interactions of mesonic bound states of the messenger
quarks\footnote{Although the contribution to $m_{1/2}$ from any single
  meson is 't~Hooft suppressed, the sum of all contributions could be
  comparable to~\eqref{eq:gauginomass}.}.  The fact that meson
messengers are weakly coupled in the 't~Hooft limit required the
authors of~\cite{Benini:2009ff} to use a combination of perturbative
field theory and holographic techniques.  In contrast, the reduced
amount of $R$-symmetry allowed us to compute the leading order
contribution to the gaugino mass using only holography as the
effective degrees of freedom are strongly coupled messengers.

Identification of the SUSY-breaking state in the dual gauge theory
relies on the large radius behavior of the bulk gravitational fields.
Since the solution used here is a small $\tau$ expansion, it is not
useful for such an analysis.  However, even using the large radius
solution, it is not clear how to make the identification of the state
in terms of dominant $F$-term or $D$-term
breaking~\cite{DeWolfe:2008zy} and indeed at strong coupling the
distinction may not be sharp (though the weakly coupled mesonic states
discussed in~\cite{Benini:2009ff} do feel effective $F$-terms).
However, to leading order in $\tau_{\mathrm{min}}$, the
contribution~\eqref{eq:gauginomass} can expressed as
\begin{equation}
  \label{eq:gauginomassb}
  \delta m_{1/2}\sim g_{\mathrm{vis}}^{2}
  \lambda\left(\Lambda_{\ve}\right)
  \frac{\Lambda_{\s}^{4}}{m_{\chi}^{3}}
  \left(\left(\frac{m_{\chi}}{\Lambda_{\ve}}\right)^{3/2}-1\right)^{3/2}.
\end{equation}
This suggests that there is some $F$-term component to the
SUSY-breaking breaking given by
\begin{equation}
  F=\sqrt{\lambda\bigl(\Lambda_{\ve}\bigr)}\Lambda_{\s}^{2}.
\end{equation}
The contribution to the gaugino mass from physics above $m_{\chi}$ is then
\begin{equation}
  \delta m_{1/2}\sim g_{\mathrm{vis}}^{2}\frac{F^{2}}{m_{\chi}^{3}}
  \left(\left(\frac{m_{\chi}}{\Lambda_{\ve}}\right)^{3/2}-1\right)^{3/2}.
\end{equation}
The fact that the gaugino mass occurs at higher order in $F$ is
similar to other examples of semi-direct gauge mediation where
$m_{1/2}$ vanishes to leading order in
$F$~\cite{Seiberg:2008qj,Benini:2009ff,Ibe:2009bh}.

This contribution to the gaugino mass na\"ively vanishes for
$m_{\chi}=\Lambda_{\ve}$. We emphasize however that at this point the
supergravity description breaks down since the $\uD 7$s reach the
curvature singularity where stringy effects are important and there
may be important corrections to~\eqref{eq:gauginomassb}.  However, the
fact that it decreases with $\tau_{\mathrm{min}}$ may not be
surprising.  The statement that the integral is dominated near
$\tau_{\mathrm{min}}$ corresponds to the statement on the field theory
side that the dominant contribution to the gaugino mass is from
physics near $m_{\chi}$.  Decreasing $\tau_{\mathrm{min}}$ corresponds
to taking $m_{\chi}$ closer to $\Lambda_{\ve}$ and so for smaller
$\tau_{\mathrm{min}}$, the integral is dominated by physics at lower
scales.  Since the effective 't~Hooft coupling decreases as the scale
decreases, heuristically one might expect that this contribution to
the gaugino mass also decreases.

The contribution~\eqref{eq:gauginomass} is a result of a calculation
of a soft SUSY-breaking term that, unlike many other examples, at no
point required the assumption of weak coupling in messenger or visible
sectors aside from the gauge coupling to the standard model.  Because
the solutions presented in~\cite{MSS1} were given only as a power
series in $\tau$, it was not possible to find an exact expression for
the gaugino mass.  However, we emphasize that this difficulty is a
very distinct difficulty from that usually faced by strong coupling in
that one could in principle use the gauge-gravity correspondence to
find an exact result in the limit of large 't~Hooft coupling.  In
contrast, without holography it is not clear how even to perform this
calculation, even in principle.

The messenger mass parameter $\mu$ has unit $R$-charge, so one would
expect that for $\mu\neq 0$, even in the regime
$m_{\chi}\gg\Lambda_{\ve}$, there would be a contribution to $m_{1/2}$
from physics at all scales.  However, for energies above $m_{\chi}$,
the messenger quark $\chi$ is no longer integrated out of the
effective field theory and the $R$-symmetry breaking effects are
suppressed such that contributions to the gaugino mass occur only at
subleading order in $\Lambda_{\s}$.  At least as far as the gaugino
mass is concerned, the $R$-symmetry breaking effect of a
non-vanishing $\mu$ is less important than the $R$-symmetry breaking
effect of confinement, though it would be worthwhile to develop a
clearer picture.  One possible step in this direction would be to take
into account the back reaction of the $\uD 7$-branes, essentially
moving away from the quenched approximation in the field theory.  For
general $\mu$, such a back reaction would break the symmetry of the
solution that is dual to the field theory $R$-symmetry.

It is well known that for theories of semi-direct or direct gauge
mediation in which the hidden sector has large rank, one typically
runs into a problem of visible sector Landau poles.  Even in the
regime of $m_{\chi}\gg\Lambda_{\ve}$ discussed
in~\cite{Benini:2009ff}, avoidance of visible sector Landau poles
forced $m_{\chi}$ to be large.  However, it was also suggested
in~\cite{Benini:2009ff} that the problem may be avoided by orbifolding
the geometry.  Although such a method might be needed to achieve
realistic soft terms, we defer such analysis to future work.

An additional interesting future direction would be to holographically
realize the visible sector matter fields as well.  In the model
of~\cite{Benini:2009ff}, the matter fields are taken to be elementary
fields, living on the UV cutoff with other soft terms resulting from
gaugino mediation.  A more complete holographic realization of gauge
mediation could involve a more detailed model in which the
supersymmetric standard model (or some extension) is realized on a
network of intersecting $\uD 7$ branes carrying non-vanishing
world-volume flux.

\acknowledgments We would like to thank H.-Y.~Chen, G.~Kane,
I.-W.~Kim, F.~Marchesano, P.~Ouyang, X.~Tata, A.~Uranga, and
L.-T.~Wang for discussions and comments.  PM and GS also thank T.~Liu
for collaboration and preliminary discussions on this and related
topics. We would like to thank the Institute for Advanced Study and
the Hong Kong Institute for Advanced Study, Hong Kong University of
Science and Technology for hospitality and support. PM and GS also
thank the Standford Institute for Theoretical Physics and SLAC for
hospitality while some preliminary discussions were held.  YS was
supported by the Nishina Memorial Foundation.  PM and GS were
supported in part by NSF CAREER Award No. Phy-0348093, DOE grant
DE-FG-02-95ER40896, a Cottrell Scholar Award from Research
Corporation, a Vilas Associate Award from the University of Wisconsin,
and a John Simon Guggenheim Memorial Foundation Fellowship.  GS also
would like to acknowledge support from the Ambrose Monell Foundation
during his stay at the Institute for Advanced Study.

\newpage
\appendix

\section{Conventions}
\label{app:conv}

Our index notation is summarized in Table~\ref{table:index}.
\TABLE[t]{
  \begin{tabular}{|c|c|}
    \hline
    label & direction \\
    \hline\hline
    $M,N$ & any direction \\
    $\mu,\nu$ & 4D Minkowski \\
    $m,n$ & radial and internal angular \\
    $\alpha,\beta$ & $\uD 7$ world-volume coordinate \\
    $a,b$ & $\uD 7$ world-volume coordinate: radial or internal angular \\
    $i,j$ & complex coordinate $z_{i}$ defined in~\eqref{eq:defconifold} \\
    $I,J$ & complex coordinates $Z_{I}$ defined in~\eqref{eq:Zcoordinates} \\
    $\sigma,\rho$ & complex $\uD 7$ world-volume coordinates \\
    \hline
  \end{tabular}
  \caption{\label{table:index}Index conventions.  Exceptions should be
    clear from context.  An index with a tilde (e.g. $\tilde{\alpha}$)
    indicates an index raised with the unwarped metric.}
}

We work in the type IIB supergravity limit of string theory where the
low energy effective action in the 10D Einstein frame
is~\cite{Polchinski:2000uf}
\begin{subequations}
  \begin{align}
    S_{\mathrm{IIB}}=&S_{\mathrm{NS}}+S_{\mathrm{R}}+S_{\mathrm{CS}}, \\
    S_{\mathrm{IIB}}^{\mathrm{NS}}=&\frac{1}{2\kappa_{10}^{2}}
    \int\ud^{10}x\, \sqrt{-\det\left(g\right)}
    \biggl[R-\frac{1}{2}\partial_{M}\Phi\partial^{M}\Phi
    -\frac{g_{s}}{2\cdot 3!}\ue^{-\Phi}
    H_{3}^{2}\biggr], \\
    S_{\mathrm{IIB}}^{\mathrm{R}}=&-\frac{1}{4\kappa_{10}^{2}}
    \int\ud^{10}x\, \sqrt{-\det\left(g\right)}
    \biggl[\ue^{2\Phi}\partial_{M}C\partial^{M}C
    +\frac{g_{s}\ue^{\Phi}}{3!}  \tilde{F}_{3}^{2}
    +\frac{g_{s}^{2}}{2\cdot 5!}\tilde{F}_{5}^{2}\biggr], \\
    S_{\mathrm{IIB}}^{\mathrm{CS}}=&\frac{g_{s}^{2}}{4\kappa_{10}^{2}}\int
    C_{4}\wedge H_{3}\wedge F_{3},
  \end{align}
\end{subequations}
where in terms of the R-R potentials $C$, $C_{2}$, and $C_{4}$ and the NS-NS
potential $B_{2}$,
\begin{equation}
  \tilde{F}_{3} = \ud C_{2} - C H_{3}, \qquad 
  \tilde{F}_{5} = \ud C_{4} + B_{2} \wedge \ud C_{2}.
\end{equation}
$R$ is the Ricci scalar built from the metric $g$ and $\Phi$ is the
dilaton such that $\bigl<\Phi\bigr>=\log g_{s}$.  The self-duality of the
5-form field strength is imposed at the level of the equations of
motion and we write
\begin{equation}
  F_{5}=\bigl(1+\ast_{10}\bigr)\mathcal{F}_{5},
\end{equation}
where $\ast_{10}$ is the Hodge star built from $g$.  The gravitational
coupling is
$2\kappa_{10}^{2}=\left(2\pi\right)^{7}\alpha'^{4}g_{s}^{2}$ and the
Einstein frame metric $g_{MN}$ is related to the string frame metric
$g^{s}_{MN}$ by the Weyl transformation
$g_{MN}=g_{s}^{1/2}\ue^{-\Phi/2}g^{s}_{MN}$.

In the Einstein frame, the action for a $\uD p$ brane is
\begin{equation}
  S_{\uD p}=S_{\uD p}^{\mathrm{DBI}}+S_{\uD p}^{\mathrm{CS}}
  + S_{\uD p}^{\mathrm{F}},
\end{equation}
where the bosonic part is,
\begin{align}
  S_{\uD p}^{\mathrm{DBI}}=& -\tau_{\uD
    p}g_{s}^{-\left(\frac{p-3}{4}\right)\Phi} \int\ud^{p+1}\xi\,
  \ue^{-\frac{\left(p-3\right)}{4}}
  \sqrt{\abs{\det\bigl(\M_{\alpha\beta}\bigr)}}, \\
  S_{\uD p}^{\mathrm{CS}}=& \tau_{\uD p}\int\mathrm{P}\biggl[
  \sum_{n}C_{n}\wedge\ue^{-B_{2}}\biggr]\wedge \ue^{2\pi\alpha'
    f_{2}},
\end{align}
with
\begin{equation}
  \M_{\alpha\beta}=
  \gamma_{\alpha\beta}+
  g_{s}^{1/2}\ue^{-\Phi/2}b_{\alpha\beta}+
  g_{s}^{1/2}\ue^{-\Phi/2}\left(2\pi\alpha'\right)f_{\alpha\beta},
\end{equation}
where $\gamma_{\alpha\beta}$ and $b_{\alpha\beta}$ are the pullbacks
of the metric and the NS-NS 2-form potential onto the world-volume of
the brane, $f_{2}=\ud A_{1}$ is the field strength for the vector
potential living on the world-volume, $\mathrm{P}\bigl[\cdot\bigr]$
indicates a pullback, and the tension of a $\uD p$-brane satisfies
$\tau_{\uD
  p}^{-1}=\left(2\pi\right)^{p}\alpha'^{\left(p+1\right)/4}g_{s}^{-1}$.
$\overline{\uD p}$-branes are distinguished by an overall sign in
front of the Chern-Simons piece.  Although the generalization to the
non-Abelian case is known~\cite{Myers:1999ps}, it is not needed for
our purposes.

The fermionic action for a $\uD p$ brane can be expanded out to
quadratic order to give a Dirac-like action~\cite{Martucci:2005rb}.
In the Einstein frame, this is given by\footnote{The conventions here
  differ from those
  in~\cite{Martucci:2005rb,Marchesano:2008rg} by an opposite sign
  for $H_{3}$.}
\begin{equation}
  S_{\uD p}^{\left(\mathrm{F}\right)}
  =\ui\tau_{\uD p}g_{s}^{-\left(\frac{p-3}{4}\right)}
  \int\ud^{8}\xi\, \ue^{\left(\frac{p-3}{4}\right)\Phi}
  \sqrt{\det\bigl(\M\bigr)}
  \overline{\Theta} P_{-}^{\uD p}\bigl[
  \bigl(\mathcal{\M}^{-1}\bigr)^{\alpha\beta}
  \Gamma_{\beta}\bigl(\mathcal{D}_{\alpha}
  +\frac{1}{4}\Gamma_{\alpha}\mathcal{O}\bigr)
  -\mathcal{O}\bigr]\Theta,
\end{equation}
where $\mathcal{D}_{\alpha}$ and $\mathcal{O}$ are the pullbacks of
operators involved in the Einstein-frame SUSY transformations of the
gravitino and dilatino,
\begin{equation}
  \delta\Psi_{M}=\mathcal{D}_{M}\epsilon,\qquad
  \delta\lambda_{\Phi}=\mathcal{O}\epsilon,
\end{equation}
where these are related to string frame fields by
\begin{equation}
  \epsilon=g_{s}^{1/8}\ue^{-\Phi/8}\epsilon^{s},\qquad
  \lambda_{\Phi}=g_{s}^{-1/8}\ue^{\Phi/8}\lambda^{s},\qquad
  \Psi_{M}=g_{s}^{1/8}\ue^{-\Phi/8}\bigl(\Psi^{s}_{M}
  -\frac{1}{4}\Gamma_{M}^{\mathrm{s}}\lambda_{\Phi}^{s}\bigr),\qquad
\end{equation}
and similarly $\Theta=g_{s}^{1/8}\ue^{-\Phi/8}\Theta^{s}$.
$\Theta=\left(\theta_{1}\ \theta_{2}\right)^{\mathrm{T}}$ is a doublet
of 10D Majorana-Weyl spinors satisfying
$\Gamma_{\left(10\right)}\theta_{i}=\theta_{i}$ where
$\Gamma_{\left(10\right)}$ is the 10D chirality operator.
$\bar{\Theta}$ is defined by
\begin{equation}
  \bar{\Theta}=\begin{pmatrix}
    \bar{\theta}_{1} & \bar{\theta}_{2}\end{pmatrix}.
\end{equation}
Following~\cite{Martucci:2005rb}, we take the $\Gamma$-matrices to be
real implying $\bar{\theta}=\theta^{T}\Gamma^{\underline{0}}$ where
underlined indices denote ``flat'' $\Gamma$-matrices.
$P_{-}^{\uD p}$ is a projection operator defined by
\begin{equation}
  P_{\pm}^{\uD p}=\frac{1}{2}\bigl(1\pm\Gamma_{\uD p}\bigr)
  =\frac{1}{2}\begin{pmatrix}
    1 & \pm\breve{\Gamma}_{\uD p}^{-1} \\ \pm\breve{\Gamma}_{\uD p} & 1
  \end{pmatrix},
\end{equation}
with
\begin{equation}
  \breve{\Gamma}_{\uD p}=
  \ui^{\left(p-2\right)\left(p-3\right)}
  \Gamma_{\uD p}^{\left(0\right)}\Lambda\left(\mathcal{F}\right),
\end{equation}
where $\Gamma_{\uD p}^{\left(0\right)}$ is given by
\begin{equation}
  \Gamma_{\uD p}^{\left(0\right)}
  =\frac{1}{\left(p+1\right)!\sqrt{-\det\left(\gamma\right)}}
  \epsilon_{\alpha_{1}\ldots\alpha_{p+1}}
  \Gamma^{\alpha_{1}\ldots\alpha_{p+1}},
\end{equation}
and
\begin{equation}
  \Lambda\left(\mathcal{F}\right)=
  \frac{\sqrt{\det\left(\gamma\right)}}
  {\sqrt{\det\left(M\right)}}
  \sum_{q}\frac{\bigl(g_{s}\ue^{-\Phi/2}\bigr)^{q/2}}{q!2^{q}}
  \mathcal{F}_{\alpha_{1}\alpha_{2}}\cdots\mathcal{F}_{a_{2q-1}\alpha_{2q}}
  \Gamma^{\alpha_{1}\cdots\alpha_{2q}},
\end{equation}
where $\mathcal{F}_{2}=b_{2}+2\pi\alpha'f_{2}$.  In terms of the usual
$\left(2k+2\right)$-dimensional chirality matrix
\begin{equation}
  \Gamma_{\left(2k+2\right)}=\ui^{k}\Gamma^{\underline{0\cdots 2k+1}},
\end{equation}
where $\Gamma^{\underline{M}}$ is a flat $\Gamma$-matrix, we have
\begin{equation}
  \ui^{\left(p-2\right)\left(p-3\right)}\Gamma_{\uD p}^{\left(0\right)}
  =\ui^{\left(p-1\right)/2}\Gamma_{\left(p+1\right)}.
\end{equation}
Except for $\Gamma_{\uD p}$, all $\Gamma$-matrices act as identity on
the doublet space,
\begin{equation}
  \Gamma_{M}\Theta=\begin{pmatrix}
    \Gamma_{M}\theta_{1} \\ \Gamma_{M}\theta_{2}\end{pmatrix}.
\end{equation}

In IIB,
\begin{equation}
  \hat{\M}_{\alpha\beta}
  =\gamma_{\alpha\beta}+g_{s}^{1/2}\ue^{-\Phi/2}
  \mathcal{F}_{\alpha\beta}\Gamma_{\left(10\right)}
  \otimes\sigma_{3},
\end{equation}
where the Pauli matrices act on the doublet space so that we can
effectively write
\begin{equation}
  \overline{\Theta}\hat{\M}_{\alpha\beta}
  =\overline{\Theta}\begin{pmatrix} \M_{\beta\alpha} & \\ & \M_{\alpha\beta}
  \end{pmatrix}.
\end{equation}

In the IIB Einstein frame,
\begin{subequations}
  \begin{align}
    \mathcal{O}=&\frac{1}{2}\Gamma^{M}\partial_{M}\Phi
    -\frac{1}{2}\ue^{\Phi}\Gamma^{M}\partial_{M}C\bigl(\ui\sigma_{2}\bigr)
    -\frac{1}{4}g_{s}^{1/2}\ue^{\Phi/2}\mathcal{G}_{3}^{+}, \\
    \mathcal{D}_{M}=& \nabla_{M}
    +\frac{1}{4}\ue^{\Phi}\partial_{M}C\left(\ui\sigma_{2}\right)
    +\frac{1}{8}\ue^{\Phi/2}g_{s}^{1/2}
    \bigl(\mathcal{G}_{3}^{-}\Gamma_{M}+
    \frac{1}{2}\Gamma_{M}\mathcal{G}_{3}^{-}\bigr) \notag \\
    &\qquad+\frac{1}{16\cdot 5!}
    g_{s}\tilde{F}_{NPQRT}\Gamma^{NPQRT}\Gamma_{M} \bigl(\ui\sigma_{2}\bigr),
  \end{align}
\end{subequations}
where
\begin{equation}
  \mathcal{G}_{3}^{\pm}
  =\frac{1}{3!}\bigl(\tilde{F}_{MNP}\sigma_{1}\pm\ue^{-\Phi}H_{MNP}\sigma_{3}\bigr)
  \Gamma^{MNP}.
\end{equation}
$\sigma_{i=1,2,3}$ are the usual Pauli matrices
\begin{equation}
  \sigma_{1}=\begin{pmatrix} 0 & 1 \\ 1 & 0\end{pmatrix},\qquad
  \sigma_{2}=\begin{pmatrix} 0 & -\ui \\ \ui & 0\end{pmatrix},\qquad
  \sigma_{3}=\begin{pmatrix} 1 & 0 \\ 0 & -1\end{pmatrix}.
\end{equation}

The non-Abelian generalization of the action in~\cite{Martucci:2005rb}
is not known.  However, to leading order in $\alpha'$, and as long as
the transverse fluctuations are suppressed, the non-Abelian action
should result from promoting $\theta$ to an adjoint-valued field and
the regular derivative to a gauge covariant derivative, and tracing
over gauge indices.

\section{Deformed conifold geometry}
\label{app:conifold}

Here we briefly review the geometry of the conifold and its
deformation following closely the discussion in~\cite{Herzog:2001xk}
(though see also~\cite{Candelas:1989js}).  The deformed conifold can
be described as the locus of points satisfying
\begin{equation}
  \label{eq:defconifold}
  \sum_{i=1}^{4}z_{i}^{2}=\ve^{2},
\end{equation}
and the singular conifold is recovered for $\ve=0$.
Eq.~\eqref{eq:defconifold} is invariant under the $\mathbb{Z}_{2}$
transformation $z_{i}\to -z_{i}$ and the $\SO{4}$ transformation
$z_{i}\to O_{ij} z_{j}$.  The radial coordinates $\tau$ and $r$ are
defined by
\begin{equation}
\label{eq:radialcoordinates}
  z_{i}\bar{z}_{i}=\ve^{2}\cosh\tau=r^{3}.
\end{equation}
The angular space is an $S^{3}$ fibered over an $S^{2}$ and is
frequently written in terms of angular coordinates
$\theta_{i=1,2}\in\left[0,\pi\right]$,
$\phi_{i=1,2}\in\left[0,2\pi\right)$, and $\psi\in\left[0,4\pi\right)$
related to the $z_{i}$ by
\begin{align}
  \frac{z_{1}}{\ve}=& \cosh\biggl(\frac{S}{2}\biggr)
  \cos\biggl(\frac{\theta_{1}+\theta_{2}}{2}\biggr)
  \cos\biggl(\frac{\phi_{1}+\phi_{2}}{2}\biggr)
  +\ui\sinh\biggl(\frac{S}{2}\biggr)
  \cos\biggl(\frac{\theta_{1}-\theta_{2}}{2}\biggr)
  \sin\biggl(\frac{\phi_{1}+\phi_{2}}{2}\biggr),  \notag \\
  \frac{z_{2}}{\ve}=& -\cosh\biggl(\frac{S}{2}\biggr)
  \cos\biggl(\frac{\theta_{1}+\theta_{2}}{2}\biggr)
  \sin\biggl(\frac{\phi_{1}+\phi_{2}}{2}\biggr)
  +\ui\sinh\biggl(\frac{S}{2}\biggr)
  \cos\biggl(\frac{\theta_{1}-\theta_{2}}{2}\biggr)
  \cos\biggl(\frac{\phi_{1}+\phi_{2}}{2}\biggr),  \notag \\
  \frac{z_{3}}{\ve}=& -\cosh\biggl(\frac{S}{2}\biggr)
  \sin\biggl(\frac{\theta_{1}+\theta_{2}}{2}\biggr)
  \cos\biggl(\frac{\phi_{1}-\phi_{2}}{2}\biggr)
  +\ui\sinh\biggl(\frac{S}{2}\biggr)
  \sin\biggl(\frac{\theta_{1}-\theta_{2}}{2}\biggr)
  \sin\biggl(\frac{\phi_{1}-\phi_{2}}{2}\biggr),  \notag \\
  \frac{z_{4}}{\ve}=& -\cosh\biggl(\frac{S}{2}\biggr)
  \sin\biggl(\frac{\theta_{1}+\theta_{2}}{2}\biggr)
  \sin\biggl(\frac{\phi_{1}-\phi_{2}}{2}\biggr)
  -\ui\sinh\biggl(\frac{S}{2}\biggr)
  \sin\biggl(\frac{\theta_{1}-\theta_{2}}{2}\biggr)
  \cos\biggl(\frac{\phi_{1}-\phi_{2}}{2}\biggr),
\end{align}
with $S=\tau+\ui\psi$.  The $\mathbb{Z}_{2}$ symmetry is then realized
as $\psi\to \psi+2\pi$.  It is convenient to define,
\begin{subequations}
  \begin{align}
    e_{1}=&-\sin\theta_{1}\ud\phi_{1}, \\
    e_{2}=&\ud\theta_{1},\\
    e_{3}=&\cos\psi\sin\theta_{2}\ud\phi_{2}-\sin\psi\ud\theta_{2},\\
    e_{4}=&\sin\psi\sin\theta_{2}\ud\phi_{2}+\cos\psi\ud\theta_{2},\\
    e_{5}=&\ud\psi+\cos\theta_{1}\ud\phi_{1}+\cos\theta_{2}\ud\phi_{2}.
  \end{align}
\end{subequations}
The metric for the deformed conifold is diagonal in the basis of
$1$-forms given by~\cite{Minasian:1999tt}
\begin{subequations}
  \begin{align}
    g_{1}=&\frac{1}{\sqrt{2}}\left(e_{1}-e_{3}\right), \\
    g_{2}=&\frac{1}{\sqrt{2}}\left(e_{2}-e_{4}\right), \\
    g_{3}=&\frac{1}{\sqrt{2}}\left(e_{1}+e_{3}\right), \\
    g_{4}=&\frac{1}{\sqrt{2}}\left(e_{2}+e_{4}\right), \\
    g_{5}=&e_{5}.
  \end{align}
\end{subequations}
In terms of the complex coordinates, there are relatively simple
expressions available for the $\SO{4}$ invariant 1-forms
\begin{equation}
    \ud\tau = \frac{1}{\ve^2\sinh\tau} \bigl(z_{i}\ud\bar{z}_{i} +
    \bar{z}_{i}\ud z_{i}\bigr),\qquad
    g_{5} = \frac{\ui}{\ve^2\sinh\tau}\bigl(z_{i}\ud\bar{z}_{i} -
    \bar{z}_{i}\ud z_{i}\bigr).
\end{equation}
Similarly, for the $\SO{4}$-invariant $2$-forms
\begin{subequations}
  \begin{align}
    g_{1}\wedge g_{2}=& \frac{\ui\bigl(1+\cosh\tau)}
    {2\ve^{4}\sinh^{3}\tau}\epsilon_{ijkl}\bigl( 2z_{i}\bar{z}_{j}\ud
    z_{k}\wedge\ud\bar{z}_{l}- z_{i}\bar{z}_{j}\ud z_{k}\wedge\ud
    z_{l}-
    z_{i}\bar{z}_{j}\ud\bar{z}_{k}\wedge\ud\bar{z}_{l} \bigr),\\
    g_{3}\wedge g_{4}=&\frac{\ui\tanh\left(\frac{\tau}{2}\right)}
    {2\ve^{4}\sinh^{2}\tau}\epsilon_{ijkl}\bigl( 2z_{i}\bar{z}_{j}\ud
    z_{k}\wedge\ud\bar{z}_{l}+ z_{i}\bar{z}_{j}\ud z_{k}\wedge \ud
    z_{l} +
    z_{i}\bar{z}_{j}\ud\bar{z}_{k}\wedge\ud\bar{z}_{l} \bigr),\\
    g_{1}\wedge g_{3} + g_{2} \wedge g_{4} =&
    \frac{1}{\ve^{4}\sinh^{2}\tau}\epsilon_{ijkl}\bigl(
    -z_{i}\bar{z}_{j}\ud z_{k}\wedge\ud z_{l} +
    z_{i} \bar{z}_{j} \ud\bar{z}_{k} \wedge \ud\bar{z}_{l}\bigr),\\
    g_{2}\wedge g_{3} + g_{4}\wedge g_{1} =&
    -\frac{2\ui\cosh\tau}{\ve^{4}\sinh^3\tau} \bigl(\bar{z}_{j}\ud
    z_{j}\bigr)\wedge \bigl(z_{i}\ud\bar{z}_{i}\bigr) +
    \frac{2\ui}{\ve^2\sinh\tau}\ud z_{i} \wedge \ud\bar{z}_i.
  \end{align}
\end{subequations}
The other 1-forms $g_{i}$ do not seem to be as easily expressed in
terms of the holomorphic coordinates.  However, one can show that
\begin{subequations}
  \begin{align}
    g_{1}^{2}+g_{2}^{2}=&-\frac{1}
    {2\ve^{4}\sinh^{2}\left(\tau/2\right)\sinh^{2}\tau} \biggl[
    \bigl(\bar{z}_{i}\ud z_{i}\bigr)^{2}+
    \bigl(z_{i}\ud\bar{z}_{i}\bigr)^{2}+
    2\cosh\tau\bigl(\bar{z}_{i}\ud z_{i}\bigr)
    \bigl(z_{i}\ud\bar{z}_{i}\bigr)\biggr. \notag \\
    &\biggl.\phantom{-\frac{1}
      {2\ve^{4}\sinh^{2}\left(\tau/2\right)\sinh^{2}\tau}\biggl[}+
    \ve^{2}\sinh^{2}\tau \bigl(\ud z_{i}\ud z_{i}
    +\ud\bar{z}_{i}\ud\bar{z}_{i}
    -2\ud z_{i}\ud\bar{z}_{i}\bigr)\biggr], \\
    g_{3}^{2}+g_{4}^{2}=&\frac{1}
    {2\ve^{4}\cosh^{2}\left(\tau/2\right)\sinh^{2}\tau} \biggl[
    \bigl(\bar{z}_{i}\ud z_{i}\bigr)^{2}+
    \bigl(z_{i}\ud\bar{z}_{i}\bigr)^{2}-
    2\cosh\tau\bigl(\bar{z}_{i}\ud z_{i}\bigr)
    \bigl(z_{i}\ud\bar{z}_{i}\bigr)\biggr. \notag \\
    &\phantom{\frac{1}
      {2\ve^{4}\cosh^{2}\left(\tau/2\right)\sinh^{2}\tau}\biggl[}
    \biggl.+\ve^{2}\sinh^{2}\tau \bigl(\ud z_{i}\ud z_{i}
    +\ud\bar{z}_{i}\ud\bar{z}_{i} +2\ud
    z_{i}\ud\bar{z}_{i}\bigr)\biggr].
  \end{align}
\end{subequations}

Using these expressions and the metric and fluxes given
in Section~\ref{subsec:gravsol}, the components of the $3$-form flux
having mixed holomorphic and anti-holomorphic indices (with
respect to the complex structure of the unperturbed KS solution) are
\begin{subequations}
\label{eq:generalmixed}
  \begin{align} 
    G_{3}^{\left(2,1\right)} =\frac{M\alpha'}{2\ve^{6}}
    &\left[2\left(a_{1}^{+}+a_{2}^{+}\right) \left(\bar{z}_{m}\ud
        z_{m}\right)\wedge \left(\epsilon_{ijk\ell}z_{i}\bar{z}_{j}
        \ud z_{k}\wedge\ud\bar{z}_{\ell}\right)\right. \notag \\
    & \left.+\left(a_{1}^{-}-a_{2}^{-}-a_{3}^{+}\right)
      \left(z_{m}\ud\bar{z}_{m}\right)\wedge
      \left(\epsilon_{ijk\ell}z_{i}\bar{z}_{j}
        \ud z_{k}\wedge\ud z_{\ell}\right)\right], \\
    G_{3}^{\left(1,2\right)} =\frac{M\alpha'}{2\ve^{6}}
    &\left[2\left(a_{1}^{-}+a_{2}^{-}\right) \left(z_{m}\ud
        \bar{z}_{m}\right)\wedge
      \left(\epsilon_{ijk\ell}z_{i}\bar{z}_{j}
        \ud z_{k}\wedge\ud\bar{z}_{\ell}\right)\right. \notag \\
    & \left.+\left(a_{1}^{+}-a_{2}^{+}-a_{3}^{-}\right)
      \left(\bar{z}_{m}\ud z_{m}\right)\wedge
      \left(\epsilon_{ijk\ell}z_{i}\bar{z}_{j} \ud
        \bar{z}_{k}\wedge\ud \bar{z}_{\ell}\right)\right],
  \end{align}
\end{subequations}
with
\begin{subequations}
  \begin{align}
    a_{1}^{\pm}\left(\tau\right) =&
    \frac{\tanh\frac{\tau}{2}}{2\sinh^{3}\tau}
    \left(\pm\left(1-F\right)+g_{s}\ue^{-\Phi}k'\right), \\
    a_{2}^{\pm}\left(\tau\right) =&\frac{1+\cosh\tau}{2\sinh^{4}\tau}
    \left(\pm F + g_{s}\ue^{-\Phi}f'\right), \\
    a_{3}^{\pm}\left(\tau\right) =&\frac{1}{\sinh^{3}\tau} \left(\pm
      F'+g_{s}\ue^{-\Phi}\frac{k-f}{2}\right).
  \end{align}
\end{subequations}
The components with pure holomorphic and pure anti-holomorphic indices
are
\begin{subequations}
\label{eq:generalpure}
  \begin{align}
    \label{eq:general30component}
    G_{3}^{\left(3,0\right)}&= \frac{M\alpha'}{2\ve^{6}}
    \left[\left(\left(1-F\right)
        \frac{\tanh\frac{\tau}{2}}{2\sinh^{3}\tau}
        -F\frac{1+\cosh\tau}{2\sinh^{4}\tau}
        -\frac{F'}{\sinh^{3}\tau}\right)\right. \notag \\
    &\left.+g_{s}\ue^{-\Phi}\left(
        -f'\frac{1+\cosh\tau}{2\sinh^{4}\tau}
        +k'\frac{\tanh\frac{\tau}{2}}{2\sinh^{3}\tau}
        +\frac{k-f}{2\sinh^{3}\tau}\right)\right] \left(\bar{z}_{m}\ud
      z_{m}\right)\wedge \left(\epsilon_{ijk\ell}z_{i}\bar{z}_{j}
      \ud z_{k}\wedge\ud z_{\ell}\right), \\
    \label{eq:general03component}
    G_{3}^{\left(0,3\right)}&= \frac{M\alpha'}{2\ve^{6}}
    \left[-\left(\left(1-F\right)
        \frac{\tanh\frac{\tau}{2}}{2\sinh^{3}\tau}
        -F\frac{1+\cosh\tau}{2\sinh^{4}\tau}
        -\frac{F'}{\sinh^{3}\tau}\right)\right. \notag \\
    &\left.+g_{s}\ue^{-\Phi}\left(
        -f'\frac{1+\cosh\tau}{2\sinh^{4}\tau}
        +k'\frac{\tanh\frac{\tau}{2}}{2\sinh^{3}\tau}
        +\frac{k-f}{2\sinh^{3}\tau}\right)\right] \left(z_{m}\ud
      \bar{z}_{m}\right)\wedge
    \left(\epsilon_{ijk\ell}z_{i}\bar{z}_{j} \ud \bar{z}_{k}\wedge\ud
      \bar{z}_{\ell}\right).
  \end{align}
\end{subequations}

For general functions $p$, $b$, $q$, and $s$, the metric will no longer be
Hermitian with respect to the complex structure of the deformed conifold.
In general, the unwarped 6D metric takes the form
\begin{align}
  \bigl(\ve^{4}\sinh^{2}\tau\bigr) \ud\tilde{s}_{6}^{2}&=
  \biggl\{p\bigl(\tau\bigr)-b\bigl(\tau\bigr)+ \frac{1}{2}\biggl[
  \frac{q\left(\tau\right)}{\cosh^{2}\left(\tau/2\right)}-
  \frac{s\left(\tau\right)}{\sinh^{2}\left(\tau/2\right)}\biggr]
  \biggr\} \bigl(\bigl(\bar{z}_{i}\ud z_{i}\bigr)^{2}+
  \bigl(z_{i}\ud\bar{z}_{i}\bigr)^{2}\bigr) \notag \\
  &+\frac{1}{2}\ve^{2}\sinh^{2}\tau
  \biggl[\frac{q\left(\tau\right)}{\cosh^{2}\left(\tau/2\right)}-
  \frac{s\left(\tau\right)}{\sinh^{2}\left(\tau/2\right)}\biggr]
  \bigl(\ud z_{i}\ud z_{i}+\ud\bar{z}_{i}\ud\bar{z}_{i}\bigr) \notag\\
  &+2\biggl\{p\bigl(\tau\bigr)+b\bigl(\tau\bigr)
  -\frac{1}{2}\cosh\bigl(\tau\bigr)\biggl[
  \frac{q\left(\tau\right)}{\cosh^{2}\left(\tau/2\right)}+
  \frac{s\left(\tau\right)}{\sinh^{2}\left(\tau/2\right)}\biggr]
  \biggr\}\bigl(\bar{z}_{i}\ud z_{i}\bigr)
  \bigl(z_{i}\ud\bar{z}_{i}\bigr) \notag \\
  & +\ve^{2}\sinh^{2}\tau\biggl[
  \frac{q\left(\tau\right)}{\cosh^{2}\left(\tau/2\right)}+
  \frac{s\left(\tau\right)}{\sinh^{2}\left(\tau/2\right)}\biggr] \ud
  z_{i}\ud\bar{z}_{i}.
  \label{eq:generalmetric}
\end{align}

The holomorphic 3-form
for the deformed conifold is
\begin{align}
  \label{eq:3form}
  \Omega=&\frac{\ve^{2}}{16\sqrt{3}}\bigl[ -\sinh\tau\bigl(g_{1}\wedge
  g_{3}+g_{2}\wedge g_{4}\bigr)+\ui\cosh\tau
  \bigl(g_{1}\wedge g_{2}-g_{3}\wedge g_{4}\bigr)\notag \\
  & \phantom{\frac{\ve^{2}}{16\sqrt{3}}\bigl[}- \ui\bigl(g_{1}\wedge
  g_{2}+g_{3}\wedge g_{4}\bigr)\bigr]
  \wedge\bigl(\ud\tau+\ui g_{5}\bigr) \notag \\
  =& \frac{1}{4\sqrt{3}\ve^{4}\sinh^{2}\tau}
  \bigl(\epsilon_{ijkl}z_{i}\bar{z}_{j}\ud z_{k}\wedge\ud z_{l}\bigr)
  \wedge\bigl(\bar{z}_{m}\ud z_{m}\bigr).
\end{align}

It also convenient to introduce another set of holomorphic
1-forms
\begin{subequations}
\label{eq:Zcoordinates}
\begin{align}
  \ud Z_{1}=&\ud\tau+\ui g_{5},\\
  \ud Z_{2}=&g_{1}-\ui\coth\frac{\tau}{2}g_{4},\\
  \ud Z_{3}=&g_{3}-\ui\tanh\frac{\tau}{2}g_{2}.
\end{align}
\end{subequations}
In these coordinates the metric~(\ref{eq:internalmetric}) is written
\begin{align}
\label{eq:internalmetricZ}
  \ud&\tilde{s}_{6}^{2}=
  \frac{1}{2}\biggl(p\bigl(\tau\bigr)+b\bigl(\tau\bigr)\biggr) \ud
  Z_{1}\ud\bar{Z}_{1} +\frac{1}{2}\biggl(s\bigl(\tau\bigr)+
  q\bigl(\tau\bigr)\tanh^{2}\frac{\tau}{2}\biggr)
  \ud Z_{2}\ud\bar{Z}_{2} \notag \\
  &+\frac{1}{2}\biggl(s\bigl(\tau\bigr)\coth^{2}\frac{\tau}{2}
  +q\bigl(\tau\bigr)\biggr)\ud Z_{3}\ud\bar{Z}_{3} \notag \\
  &+\frac{1}{4}\biggl(p\bigl(\tau\bigr)-b\bigl(\tau\bigr)\biggr)
  \bigl(\ud Z_{1}\ud Z_{1}+\ud\bar{Z}_{1}\ud\bar{Z}_{1}\bigr)
  +\frac{1}{4}\biggl(s\bigl(\tau\bigr)-
  q\bigl(\tau\bigr)\tanh^{2}\frac{\tau}{2}\biggr) \bigl(\ud Z_{2}\ud
  Z_{2}+\ud\bar{Z}_{2}\ud\bar{Z}_{2}\bigr)
  \notag \\
  &+\frac{1}{4}\biggl(-s\bigl(\tau\bigr)\coth^{2}\frac{\tau}{2}
  +q\bigl(\tau\bigr)\biggr) \bigl(\ud Z_{3}\ud
  Z_{3}+\ud\bar{Z}_{3}\ud\bar{Z}_{3}\bigr),
\end{align}
and the holomorphic 3-form of the deformed conifold takes the simple
form
\begin{equation}
  \Omega=-\frac{\ve^{2}}{16\sqrt{3}}\sinh\tau\ud Z_{1}
  \wedge\ud Z_{2}\wedge\ud Z_{3}.
\end{equation}
The components of $G_{3}$ with mixed holomorphic and anti-holomorphic
indices can then be written
\begin{subequations}
\label{eq:generalmixedZ}
  \begin{align}
    G_{3}^{\left(2,1\right)}=&-\frac{M\alpha'\sinh^{3}\tau}{8}\biggl\{
    \bigl(a_{1}^{+}+a_{2}^{+}\bigr) \bigl(\ud Z_{1}\wedge\ud
    Z_{2}\wedge\ud\bar{Z}_{3}-
    \ud Z_{1}\wedge\ud\bar{Z}_{2}\wedge\ud Z_{3}\bigr)\notag \\
    &\phantom{-\frac{M\alpha'\sin^{3}\tau}{8}\biggl\{}
    +\bigl(a_{1}^{-}-a_{2}^{-}-a_{3}^{+}\bigr)
    \ud\bar{Z}_{1}\wedge\ud Z_{2}\wedge\ud Z_{3}\biggr\},
    \label{eq:general21Z}\\
    G_{3}^{\left(1,2\right)}=&\frac{M\alpha'\sinh^{3}\tau}{8}\biggl\{
    \bigl(a_{1}^{-}+a_{2}^{-}\bigr)
    \bigl(\ud\bar{Z}_{1}\wedge\ud\bar{Z}_{2}\wedge\ud Z_{3}
    -\ud\bar{Z}_{1}\wedge\ud Z_{2}\wedge\ud\bar{Z}_{3}\bigr)
    \notag \\
    &\phantom{\frac{M\alpha'\sinh^{3}\tau}{8}\biggl\{}
    +\bigl(a_{1}^{+}-a_{2}^{+}-a_{3}^{-}\bigr) \ud
    Z_{1}\wedge\ud\bar{Z}_{2}\wedge\ud\bar{Z}_{3}\biggr\},
    \label{eq:general12Z}
  \end{align}
\end{subequations}
while the components with pure holomorphic or pure anti-holomorphic
indices are
\begin{subequations}
\label{eq:generalpureZ}
  \begin{align}
    G_{3}^{\left(3,0\right)}=&\frac{M\alpha'}{16}
    \biggl\{-\bigl(1-F\bigr)\tanh\frac{\tau}{2}
    +F\coth\frac{\tau}{2}+2F' \notag \\
    &\phantom{\frac{M\alpha'}{16}\biggl\{}
    +g_{s}\ue^{-\Phi}\bigl(f'\coth\frac{\tau}{2}
    -k'\tanh\frac{\tau}{2} -\bigl(f-k\bigr)\biggr)\biggr\}
    \ud Z_{1}\wedge\ud Z_{2}\wedge\ud Z_{3}, \\
    G_{3}^{\left(0,3\right)}=&\frac{M\alpha'}{16}
    \biggl\{-\bigl(1-F\bigr)\tanh\frac{\tau}{2}
    +F\coth\frac{\tau}{2}+2F' \notag \\
    &\phantom{\frac{M\alpha'}{16}\biggl\{}
    -g_{s}\ue^{-\Phi}\bigl(f'\coth\frac{\tau}{2}
    -k'\tanh\frac{\tau}{2} -\bigl(f-k\bigr)\biggr)\biggr\} \ud
    \bar{Z}_{1}\wedge\ud\bar{Z}_{2}\wedge\ud\bar{Z}_{3}.
  \end{align}
\end{subequations}

\newpage

\section{Pullbacks of bulk fields}
\label{app:pullbacks}

We consider a stack of $\uD 7$-branes satisfying the
Kuperstein embedding condition~\cite{Kuperstein:2004hy}
\begin{equation}
  z_{4}=\mu,
\end{equation}
where $z_{i}$ are holomorphic coordinates
satisfying~\eqref{eq:defconifold}.  For the purpose of computing
pullbacks onto the world-volumes, it is useful to adopt coordinates in
which the bulk geometry is seen as a foliation of Kuperstein divisors.
Such coordinates $\left(\rho,\chi,\bar{\chi}, \phi, \theta,\xi\right)$
were given in~\cite{Benini:2009ff}
\begin{subequations}
\label{eq:foliationcoordinates}
  \begin{align}
    z_{1}=&\ui\eta\bigl(\chi\bigr)\biggl[
    \cos\phi\cosh\biggl(\frac{\rho+\ui\xi}{2}\biggr) \cos\theta-
    \ui\sin\phi\sinh\biggl(\frac{\rho+\ui\xi}{2}\biggr)\biggr],\\
    z_{2}=&\ui\eta\bigl(\chi\bigr)\biggl[
    \sin\phi\cosh\biggl(\frac{\rho+\ui\xi}{2}\biggr) \cos\theta-
    \ui\cos\phi\sinh\biggl(\frac{\rho+\ui\xi}{2}\biggr)\biggr], \\
    z_{3}=&\ui\eta\bigl(\chi\bigr)\cosh\biggl(\frac{\rho+\ui\xi}{2}\biggr)
    \sin\theta, \\
    z_{4}=&\mu+\chi.
  \end{align}
\end{subequations}
where the $z_{i}$ still satisfy~(\ref{eq:defconifold}).
$\eta$ and $\rho$ are defined by
\begin{equation}
  \eta\bigl(\chi\bigr)=\sqrt{\bigl(\mu+\chi\bigr)^{2}-\ve^{2}},
\end{equation}
and $\rho$ is given by
\begin{equation}
  \ve^{2}\cosh\tau=\abs{\eta}^{2}\cosh\rho+\abs{\mu+\chi}^{2}.
\end{equation}
A Kuperstein embedding is then specified by the simple condition
$\chi=0$.  One can find the unwarped metric on the 4-cycle wrapped by
the $\uD 7$-branes by substituting the
coordinates~\eqref{eq:foliationcoordinates} into the
metric~\eqref{eq:generalmetric}.  The result is
\begin{equation}
  \ud \tilde{s}_{4}^{2}=v\bigl(\tau\bigr)\ud\rho^{2}
  + w\bigl(\tau\bigr)h_{3}^{2}+
  \frac{\eta^{2}s\left(\tau\right)\left(\cosh\rho + 1\right)}
  {2\ve^{2}\left(\cosh\tau-1\right)}h_{1}^{2}+
  \frac{\eta^{2}q\left(\tau\right)\left(\cosh\rho-1\right)}
  {2\ve^{2}\left(\cosh\tau+1\right)}h_{2}^{2},
\end{equation}
with
\begin{subequations}
  \begin{align}
    v\left(\tau\right)=\frac{\eta^{2}}{2\ve^{2}}
    \biggl\{\frac{2\eta^{2}p\left(\tau\right)\sinh^{2}\rho}
    {\sinh^{2}\tau} &+ q\bigl(\tau\bigr)\left[\frac{\ve^{2}
        (\cosh\rho+1)}{\cosh\tau+1}- \frac{\eta^2\sinh^{2}\rho}
      {(\cosh\tau+1)^2}\right]
    \notag \\
    & +s\bigl(\tau\bigr) \left[\frac{\varepsilon^2 (\cosh\rho-1)}
      {\cosh\tau-1}-\frac{\eta^2
        \sinh^{2}\rho}{(\cosh\tau-1)^2}\right]
    \biggr\}, \\
    w\left(\tau\right)=\frac{\eta^{2}}{2\ve^{2}}
    \biggl\{\frac{2\eta^{2}b\left(\tau\right)\sinh^{2}\rho}{\sinh^2\tau}
    &+q\bigl(\tau\bigr)\left[\frac{\varepsilon^2
        (\cosh\rho-1)}{\cosh\tau+1}- \frac{\eta^2
        \sinh^{2}\rho}{\sinh^2 \tau}\right]
    \notag \\
    &+s\bigl(\tau\bigr) \left[\frac{\varepsilon^2 (\cosh\rho+1)}
      {\cosh\tau-1}-\frac{\eta^2 \sinh^{2}\rho}{\sinh^2\tau}\right]
    \biggr\},
  \end{align}
\end{subequations}
where as in~\cite{Benini:2009ff}, the 1-forms $h_{i}$ are
\begin{align}
  h_{1}=&2\biggl(\cos\frac{\gamma}{2}\ud\theta-
  \sin\frac{\gamma}{2}\sin\theta\ud\phi\biggr),\\
  h_{2}=&2\biggl(\sin\frac{\gamma}{2}\ud\theta+
  \cos\frac{\gamma}{2}\sin\theta\ud\phi\biggr),\\
  h_{3}=&\ud\gamma-2\cos\theta\ud\phi.
\end{align}
For the KS geometry, this reduces to the metric given
in~\cite{Benini:2009ff},
\begin{equation}
  \ud \tilde{s}_{4}^{2}=
  \frac{K\bigl(\tau\bigr)\eta^{2}}{2\ve^{2/3}}
  \bigl[K_{2}\bigl(\rho\bigr)\bigl(\ud\rho^{2}
  + h_{3}^{2}\bigr)
  +\cosh^{2}\frac{\rho}{2}h_{1}^{2}
  +\sinh^{2}\frac{\rho}{2}h_{2}^{2}\bigr].
\end{equation}

The pullback of $B_{\left(2\right)}$ can also be written in these coordinates
as
\begin{align}
  b_{2}=-\frac{g_{s}M\alpha'}{2\ve^{4}}\eta^{3}\mu
  \biggl[&k\left(\tau\right)\csch\frac{\rho}{2}
  \csch^{3}\tau\sinh^{2}\rho\sinh^{2}\frac{\tau}{2}
  \, \ud\rho\wedge h_{2} \notag \\
  &+\frac{1}{2}f\left(\tau\right)
  \cosh^{3}\frac{\rho}{2}\csch^{3}\frac{\tau}{2} \sech\frac{\tau}{2}\,
  h_{1}\wedge h_{3}\biggr].
\end{align}

\bibliographystyle{jhep}

\bibliography{MSS-hgm}

\providecommand{\href}[2]{#2}\begingroup\raggedright\begin{thebibliography}{10}

\bibitem{Martin:1997ns}
S.~P. Martin, {\it A supersymmetry primer},
  \href{http://xxx.lanl.gov/abs/hep-ph/9709356}{{\tt hep-ph/9709356}}.

\bibitem{Chung:2003fi}
D.~Chung, L.~Everett, G.~Kane, S.~King, and J.~D. Lykken, {\it The soft
  supersymmetry-breaking {L}agrangian: {T}heory and applications},  {\em Phys.
  Rept.} {\bf 407} (2005) 1--203,
  [\href{http://xxx.lanl.gov/abs/hep-ph/0312378}{{\tt hep-ph/0312378}}].

\bibitem{Dine:1981za}
M.~Dine, W.~Fischler, and M.~Srednicki, {\it Supersymmetric technicolor},  {\em
  Nucl. Phys.} {\bf B189} (1981) 575--593.

\bibitem{Dimopoulos:1981au}
S.~Dimopoulos and S.~Raby, {\it Supercolor},  {\em Nucl. Phys.} {\bf B192}
  (1981) 353.

\bibitem{Dine:1981gu}
M.~Dine and W.~Fischler, {\it A phenomenological model of particle physics
  based on supersymmetry},  {\em Phys. Lett.} {\bf B110} (1982) 227.

\bibitem{Nappi:1982hm}
C.~R. Nappi and B.~A. Ovrut, {\it Supersymmetric extension of the {$\SU{3}
  \times \SU{2}\times\U{1}$} model},  {\em Phys. Lett.} {\bf B113} (1982) 175.

\bibitem{AlvarezGaume:1981wy}
L.~Alvarez-Gaume, M.~Claudson, and M.~B. Wise, {\it {Low-Energy
  Supersymmetry}},  {\em Nucl. Phys.} {\bf B207} (1982) 96.

\bibitem{Dimopoulos:1982gm}
S.~Dimopoulos and S.~Raby, {\it Geometric hierarchy},  {\em Nucl. Phys.} {\bf
  B219} (1983) 479.

\bibitem{Dine:1993yw}
M.~Dine and A.~E. Nelson, {\it Dynamical supersymmetry breaking at
  low-energies},  {\em Phys. Rev.} {\bf D48} (1993) 1277--1287,
  [\href{http://xxx.lanl.gov/abs/hep-ph/9303230}{{\tt hep-ph/9303230}}].

\bibitem{Dine:1994vc}
M.~Dine, A.~E. Nelson, and Y.~Shirman, {\it Low-energy dynamical supersymmetry
  breaking simplified},  {\em Phys. Rev.} {\bf D51} (1995) 1362--1370,
  [\href{http://xxx.lanl.gov/abs/hep-ph/9408384}{{\tt hep-ph/9408384}}].

\bibitem{Dine:1995ag}
M.~Dine, A.~E. Nelson, Y.~Nir, and Y.~Shirman, {\it New tools for low-energy
  dynamical supersymmetry breaking},  {\em Phys. Rev.} {\bf D53} (1996)
  2658--2669, [\href{http://xxx.lanl.gov/abs/hep-ph/9507378}{{\tt
  hep-ph/9507378}}].

\bibitem{Giudice:1998bp}
G.~F. Giudice and R.~Rattazzi, {\it Theories with gauge-mediated supersymmetry
  breaking},  {\em Phys. Rept.} {\bf 322} (1999) 419--499,
  [\href{http://xxx.lanl.gov/abs/hep-ph/9801271}{{\tt hep-ph/9801271}}].

\bibitem{Meade:2008wd}
P.~Meade, N.~Seiberg, and D.~Shih, {\it General gauge mediation},  {\em Prog.
  Theor. Phys. Suppl.} {\bf 177} (2009) 143--158,
  [\href{http://xxx.lanl.gov/abs/0801.3278}{{\tt arXiv:0801.3278}}].

\bibitem{Shadmi:1999jy}
Y.~Shadmi and Y.~Shirman, {\it Dynamical supersymmetry breaking},  {\em Rev.
  Mod. Phys.} {\bf 72} (2000) 25--64,
  [\href{http://xxx.lanl.gov/abs/hep-th/9907225}{{\tt hep-th/9907225}}].

\bibitem{Affleck:1984xz}
I.~Affleck, M.~Dine, and N.~Seiberg, {\it Dynamical supersymmetry breaking in
  four-dimensions and its phenomenological implications},  {\em Nucl. Phys.}
  {\bf B256} (1985) 557.

\bibitem{Poppitz:1996fw}
E.~Poppitz and S.~P. Trivedi, {\it New models of gauge and gravity mediated
  supersymmetry breaking},  {\em Phys. Rev.} {\bf D55} (1997) 5508--5519,
  [\href{http://xxx.lanl.gov/abs/hep-ph/9609529}{{\tt hep-ph/9609529}}].

\bibitem{ArkaniHamed:1997jv}
N.~Arkani-Hamed, J.~March-Russell, and H.~Murayama, {\it Building models of
  gauge-mediated supersymmetry breaking without a messenger sector},  {\em
  Nucl. Phys.} {\bf B509} (1998) 3--32,
  [\href{http://xxx.lanl.gov/abs/hep-ph/9701286}{{\tt hep-ph/9701286}}].

\bibitem{Murayama:1997pb}
H.~Murayama, {\it A model of direct gauge mediation},  {\em Phys. Rev. Lett.}
  {\bf 79} (1997) 18--21, [\href{http://xxx.lanl.gov/abs/hep-ph/9705271}{{\tt
  hep-ph/9705271}}].

\bibitem{Seiberg:2008qj}
N.~Seiberg, T.~Volansky, and B.~Wecht, {\it Semi-direct gauge mediation},  {\em
  JHEP} {\bf 11} (2008) 004, [\href{http://xxx.lanl.gov/abs/0809.4437}{{\tt
  arXiv:0809.4437}}].

\bibitem{Seiberg:1994pq}
N.~Seiberg, {\it Electric - magnetic duality in supersymmetric non-abelian
  gauge theories},  {\em Nucl. Phys.} {\bf B435} (1995) 129--146,
  [\href{http://xxx.lanl.gov/abs/hep-th/9411149}{{\tt hep-th/9411149}}].

\bibitem{Maldacena:1997re}
J.~M. Maldacena, {\it The large {$N$} limit of superconformal field theories
  and supergravity},  {\em Adv. Theor. Math. Phys.} {\bf 2} (1998) 231--252,
  [\href{http://xxx.lanl.gov/abs/hep-th/9711200}{{\tt hep-th/9711200}}].

\bibitem{Gubser:1998bc}
S.~S. Gubser, I.~R. Klebanov, and A.~M. Polyakov, {\it Gauge theory correlators
  from non-critical string theory},  {\em Phys. Lett.} {\bf B428} (1998)
  105--114, [\href{http://xxx.lanl.gov/abs/hep-th/9802109}{{\tt
  hep-th/9802109}}].

\bibitem{Witten:1998qj}
E.~Witten, {\it Anti-de {S}itter space and holography},  {\em Adv. Theor. Math.
  Phys.} {\bf 2} (1998) 253--291,
  [\href{http://xxx.lanl.gov/abs/hep-th/9802150}{{\tt hep-th/9802150}}].

\bibitem{Kachru:2009kg}
S.~Kachru, D.~Simic, and S.~P. Trivedi, {\it Stable non-supersymmetric throats
  in string theory},  \href{http://xxx.lanl.gov/abs/0905.2970}{{\tt
  arXiv:0905.2970}}.

\bibitem{Benini:2009ff}
F.~Benini, A.~Dymarsky, S.~Franco, S.~Kachru, D.~Simic, and H.~Verlinde, {\it
  Holographic gauge mediation},  \href{http://xxx.lanl.gov/abs/0903.0619}{{\tt
  arXiv:0903.0619}}.

\bibitem{MSS1}
P.~McGuirk, G.~Shiu, and Y.~Sumitomo, {\it Non-supersymmetric infrared
  perturbations to the warped deformed conifold},
  \href{http://xxx.lanl.gov/abs/0910.4581}{{\tt arXiv:0910.4581}}.

\bibitem{DeWolfe:2008zy}
O.~DeWolfe, S.~Kachru, and M.~Mulligan, {\it A gravity dual of metastable
  dynamical supersymmetry breaking},  {\em Phys. Rev.} {\bf D77} (2008) 065011,
  [\href{http://xxx.lanl.gov/abs/0801.1520}{{\tt arXiv:0801.1520}}].

\bibitem{Izawa:2008ef}
K.~I. Izawa and Y.~Nakai, {\it Strongly coupled semi-direct mediation of
  supersymmetry breaking},  \href{http://xxx.lanl.gov/abs/0812.4089}{{\tt
  arXiv:0812.4089}}.

\bibitem{Klebanov:2000nc}
I.~R. Klebanov and A.~A. Tseytlin, {\it Gravity duals of supersymmetric
  {$\SU{N}\times \SU{N+M}$} gauge theories},  {\em Nucl. Phys.} {\bf B578}
  (2000) 123--138, [\href{http://xxx.lanl.gov/abs/hep-th/0002159}{{\tt
  hep-th/0002159}}].

\bibitem{Klebanov:2000hb}
I.~R. Klebanov and M.~J. Strassler, {\it Supergravity and a confining gauge
  theory: Duality cascades and {$\chi$SB}-resolution of naked singularities},
  {\em JHEP} {\bf 08} (2000) 052,
  [\href{http://xxx.lanl.gov/abs/hep-th/0007191}{{\tt hep-th/0007191}}].

\bibitem{Kachru:2002gs}
S.~Kachru, J.~Pearson, and H.~L. Verlinde, {\it Brane/flux annihilation and the
  string dual of a non- supersymmetric field theory},  {\em JHEP} {\bf 06}
  (2002) 021, [\href{http://xxx.lanl.gov/abs/hep-th/0112197}{{\tt
  hep-th/0112197}}].

\bibitem{Karch:2002sh}
A.~Karch and E.~Katz, {\it Adding flavor to {AdS/CFT}},  {\em JHEP} {\bf 06}
  (2002) 043, [\href{http://xxx.lanl.gov/abs/hep-th/0205236}{{\tt
  hep-th/0205236}}].

\bibitem{Ouyang:2003df}
P.~Ouyang, {\it Holomorphic {D7}-branes and flavored {$\mathcal{N} = 1$} gauge
  theories},  {\em Nucl. Phys.} {\bf B699} (2004) 207--225,
  [\href{http://xxx.lanl.gov/abs/hep-th/0311084}{{\tt hep-th/0311084}}].

\bibitem{Kuperstein:2004hy}
S.~Kuperstein, {\it Meson spectroscopy from holomorphic probes on the warped
  deformed conifold},  {\em JHEP} {\bf 03} (2005) 014,
  [\href{http://xxx.lanl.gov/abs/hep-th/0411097}{{\tt hep-th/0411097}}].

\bibitem{Chen:2008jj}
H.-Y. Chen, P.~Ouyang, and G.~Shiu, {\it On supersymmetric {D}7-branes in the
  warped deformed conifold},  \href{http://xxx.lanl.gov/abs/0807.2428}{{\tt
  arXiv:0807.2428}}.

\bibitem{Herzog:2001xk}
C.~P. Herzog, I.~R. Klebanov, and P.~Ouyang, {\it Remarks on the warped
  deformed conifold},  \href{http://xxx.lanl.gov/abs/hep-th/0108101}{{\tt
  hep-th/0108101}}.

\bibitem{Klebanov:2002gr}
I.~R. Klebanov, P.~Ouyang, and E.~Witten, {\it A gravity dual of the chiral
  anomaly},  {\em Phys. Rev.} {\bf D65} (2002) 105007,
  [\href{http://xxx.lanl.gov/abs/hep-th/0202056}{{\tt hep-th/0202056}}].

\bibitem{'tHooft:1973jz}
G.~'t~Hooft, {\it A planar diagram theory for strong interactions},  {\em Nucl.
  Phys.} {\bf B72} (1974) 461.

\bibitem{'tHooft:1974hx}
G.~'t~Hooft, {\it A two-dimensional model for mesons},  {\em Nucl. Phys.} {\bf
  B75} (1974) 461.

\bibitem{Witten:1979kh}
E.~Witten, {\it Baryons in the {$1/N$} expansion},  {\em Nucl. Phys.} {\bf
  B160} (1979) 57.

\bibitem{Kaplan:1999ac}
D.~E. Kaplan, G.~D. Kribs, and M.~Schmaltz, {\it Supersymmetry breaking through
  transparent extra dimensions},  {\em Phys. Rev.} {\bf D62} (2000) 035010,
  [\href{http://xxx.lanl.gov/abs/hep-ph/9911293}{{\tt hep-ph/9911293}}].

\bibitem{Chacko:1999mi}
Z.~Chacko, M.~A. Luty, A.~E. Nelson, and E.~Ponton, {\it Gaugino mediated
  supersymmetry breaking},  {\em JHEP} {\bf 01} (2000) 003,
  [\href{http://xxx.lanl.gov/abs/hep-ph/9911323}{{\tt hep-ph/9911323}}].

\bibitem{Camara:2004jj}
P.~G. Camara, L.~E. Ibanez, and A.~M. Uranga, {\it Flux-induced {SUSY}-breaking
  soft terms on {D7-D3} brane systems},  {\em Nucl. Phys.} {\bf B708} (2005)
  268--316, [\href{http://xxx.lanl.gov/abs/hep-th/0408036}{{\tt
  hep-th/0408036}}].

\bibitem{Lust:2008zd}
D.~Lust, F.~Marchesano, L.~Martucci, and D.~Tsimpis, {\it Generalized
  non-supersymmetric flux vacua},  {\em JHEP} {\bf 11} (2008) 021,
  [\href{http://xxx.lanl.gov/abs/0807.4540}{{\tt arXiv:0807.4540}}].

\bibitem{Benini:2007gx}
F.~Benini, F.~Canoura, S.~Cremonesi, C.~Nunez, and A.~V. Ramallo, {\it
  Backreacting flavors in the {K}lebanov-{S}trassler background},  {\em JHEP}
  {\bf 09} (2007) 109, [\href{http://xxx.lanl.gov/abs/0706.1238}{{\tt
  arXiv:0706.1238}}].

\bibitem{Benini:2007kg}
F.~Benini, {\it A chiral cascade via backreacting {D7}-branes with flux},  {\em
  JHEP} {\bf 10} (2008) 051, [\href{http://xxx.lanl.gov/abs/0710.0374}{{\tt
  arXiv:0710.0374}}].

\bibitem{Martucci:2005rb}
L.~Martucci, J.~Rosseel, D.~Van~den Bleeken, and A.~Van~Proeyen, {\it Dirac
  actions for {D}-branes on backgrounds with fluxes},  {\em Class. Quant.
  Grav.} {\bf 22} (2005) 2745--2764,
  [\href{http://xxx.lanl.gov/abs/hep-th/0504041}{{\tt hep-th/0504041}}].

\bibitem{Marolf:2003ye}
D.~Marolf, L.~Martucci, and P.~J. Silva, {\it Fermions, {T}-duality and
  effective actions for {D}-branes in bosonic backgrounds},  {\em JHEP} {\bf
  04} (2003) 051, [\href{http://xxx.lanl.gov/abs/hep-th/0303209}{{\tt
  hep-th/0303209}}].

\bibitem{Marolf:2003vf}
D.~Marolf, L.~Martucci, and P.~J. Silva, {\it Actions and fermionic symmetries
  for {D}-branes in bosonic backgrounds},  {\em JHEP} {\bf 07} (2003) 019,
  [\href{http://xxx.lanl.gov/abs/hep-th/0306066}{{\tt hep-th/0306066}}].

\bibitem{Marchesano:2008rg}
F.~Marchesano, P.~McGuirk, and G.~Shiu, {\it Open string wavefunctions in
  warped compactifications},  {\em JHEP} {\bf 04} (2009) 095,
  [\href{http://xxx.lanl.gov/abs/0812.2247}{{\tt arXiv:0812.2247}}].

\bibitem{Camara:2003ku}
P.~G. Camara, L.~E. Ibanez, and A.~M. Uranga, {\it Flux-induced {SUSY}-breaking
  soft terms},  {\em Nucl. Phys.} {\bf B689} (2004) 195--242,
  [\href{http://xxx.lanl.gov/abs/hep-th/0311241}{{\tt hep-th/0311241}}].

\bibitem{Marino:1999af}
M.~Marino, R.~Minasian, G.~W. Moore, and A.~Strominger, {\it Nonlinear
  instantons from supersymmetric $p$-branes},  {\em JHEP} {\bf 01} (2000) 005,
  [\href{http://xxx.lanl.gov/abs/hep-th/9911206}{{\tt hep-th/9911206}}].

\bibitem{Gomis:2005wc}
J.~Gomis, F.~Marchesano, and D.~Mateos, {\it An open string landscape},  {\em
  JHEP} {\bf 11} (2005) 021,
  [\href{http://xxx.lanl.gov/abs/hep-th/0506179}{{\tt hep-th/0506179}}].

\bibitem{Ibe:2009bh}
M.~Ibe, K.~I. Izawa, and Y.~Nakai, {\it Studying gaugino mass in semi-direct
  gauge mediation},  {\em Phys. Rev.} {\bf D80} (2009) 035002,
  [\href{http://xxx.lanl.gov/abs/0907.2970}{{\tt arXiv:0907.2970}}].

\bibitem{Polchinski:2000uf}
J.~Polchinski and M.~J. Strassler, {\it The string dual of a confining
  four-dimensional gauge theory},
  \href{http://xxx.lanl.gov/abs/hep-th/0003136}{{\tt hep-th/0003136}}.

\bibitem{Myers:1999ps}
R.~C. Myers, {\it Dielectric-branes},  {\em JHEP} {\bf 12} (1999) 022,
  [\href{http://xxx.lanl.gov/abs/hep-th/9910053}{{\tt hep-th/9910053}}].

\bibitem{Candelas:1989js}
P.~Candelas and X.~C. de~la Ossa, {\it Comments on conifolds},  {\em Nucl.
  Phys.} {\bf B342} (1990) 246--268.

\bibitem{Minasian:1999tt}
R.~Minasian and D.~Tsimpis, {\it On the geometry of non-trivially embedded
  branes},  {\em Nucl. Phys.} {\bf B572} (2000) 499--513,
  [\href{http://xxx.lanl.gov/abs/hep-th/9911042}{{\tt hep-th/9911042}}].

\end{thebibliography}\endgroup
  
\end{document}